\documentclass[pra,twocolumn,showpacs,eqsecnum,superscriptaddress,floatfix]{revtex4}
\usepackage{graphicx}
\usepackage{verbatim}
\usepackage{color}
\usepackage{amsmath}
\usepackage[urlcolor=blue, hyperindex, colorlinks, bookmarks=true,linkcolor=black,citecolor=black]{hyperref}
\usepackage{microtype}

\newcommand{\dg}{^\dagger}

\newcommand{\bra}[1]{\langle{#1}|}
\newcommand{\ket}[1]{|{#1}\rangle}

\newcommand{\ito}{It\^o~}

\newcommand{\erf}[1]{Eq.~(\ref{#1})}

\newcommand{\erft}[2]{Eqs.~(\ref{#1}) -- (\ref{#2})}
\newcommand{\szo}{\op{\sigma}_z}

\newcommand{\ano}{\op{a}}
\newcommand{\cro}{\op{a}\dg}

\newcommand{\Mso}[2]{{\cal M}[{#1}]}
\newcommand{\Jp}{{\bar{J}}}

\newcommand{\eq}[1]{Eq.~(\ref{#1})}

\newcommand{\sz}{\op\sigma_{z}}
\newcommand{\sx}{\op\sigma_{x}}

\newcommand{\spp}{\op\sigma_{+}}
\newcommand{\smm}{\op\sigma_{-}}
\newcommand{\aop}{\op a}
\newcommand{\ad}{\op a^\dag}

\newcommand{\veps}{\varepsilon}
\newcommand{\wa}{\omega_{a}}
\newcommand{\wac}{\omega_\mathrm{ac}(t)}
\newcommand{\wrr}{\omega_{r}}
\newcommand{\Drr}{\Delta_{r}}

\newcommand{\w}[1]{\omega_{#1}}

\newcommand{\op}[1]{#1}
\newcommand{\agg}{\alpha_g}
\newcommand{\aee}{\alpha_e}
\newcommand{\dagg}{\dot\alpha_g}
\newcommand{\daee}{\dot\alpha_e}
\newcommand{\Pa}{\op{\Pi}_\alpha}
\newcommand{\dPa}{\dot{\op{\Pi}}_\alpha}
\newcommand{\Pg}{\op{\Pi}_g}
\newcommand{\Pe}{\op{\Pi}_e}
\newcommand{\D}[1]{{\cal  D}[#1]}

\newcommand{\com}[2]{\left[#1,#2\right]}
\newcommand{\qrho}{\rho}
\newcommand{\crho}{\varrho}
\newcommand{\tP}{\mathbf P}
\newcommand{\trans}[1]{^{#1}}

\begin{document}

\title{Quantum trajectory approach to circuit QED: Quantum jumps and the Zeno effect}
\date{\today}
\author{Jay Gambetta}
\affiliation{Departments of Applied Physics and Physics, Yale University, New Haven, Connecticut 06520, USA}
\affiliation{Institute for Quantum Computing and Department of Physics and Astronomy, University of Waterloo, Waterloo, Ontario N2L 3G1, Canada}
\author{Alexandre~Blais}
\affiliation{D\'epartement de Physique et Regroupement Qu\'eb\'ecois sur les Mat\'eriaux de Pointe, Universit\'e de Sherbrooke, Sherbrooke, Qu\'ebec, Canada, J1K 2R1}
\author{M. Boissonneault}
\affiliation{D\'epartement de Physique et Regroupement Qu\'eb\'ecois sur les Mat\'eriaux de Pointe, Universit\'e de Sherbrooke, Sherbrooke, Qu\'ebec, Canada, J1K 2R1}
\author{A. A. Houck}
\affiliation{Departments of Applied Physics and Physics, Yale University, New Haven, Connecticut 06520, USA}
\author{D. I. Schuster}
\affiliation{Departments of Applied Physics and Physics, Yale University, New Haven, Connecticut 06520, USA}
\author{S. M. Girvin}
\affiliation{Departments of Applied Physics and Physics, Yale University, New Haven, Connecticut 06520, USA}

\begin{abstract}
We present a theoretical study of a superconducting charge qubit dispersively coupled to a transmission line resonator.  Starting from a master equation description of this coupled system and using a polaron transformation, we obtain an exact effective master equation for the qubit.  We then use quantum trajectory theory to investigate the measurement of the qubit by continuous homodyne measurement of the resonator out-field.  Using the same porlaron transformation, a stochastic master equation for the conditional state of the qubit is obtained.  From this result, various definitions of the measurement time are studied.  Furthermore, we find that in the limit of strong homodyne measurement, typical quantum trajectories for the qubit exhibit a crossover from diffusive to jump-like behavior.  Finally, in the presence of Rabi drive on the qubit, the qubit dynamics is shown to exhibit quantum Zeno behavior.
\end{abstract}


\maketitle

\section{Introduction}

Continuous-in-time measurement theory~\cite{carmichael:1993a,gardiner:2004b}, or quantum trajectory theory, describes how an observer's state of knowledge of a quantum system (known as the conditional state) evolves given a measurement record.  Even in the absence of classical noise, such a trajectory follows a stochastic path in time, with the randomness being due to quantum uncertainty. These stochastic trajectories are either diffusive or jump-like in nature.  Diffusive trajectories usually arise when the observable being measured is only weakly coupled to the detector~\cite{wiseman:2001a}, whereas jump-like behavior occurs when there is a large sudden change in the observers knowledge of the system state, a typical example of the latter being detection of a photon with a photomultiplier~\cite{carmichael:1993a,molmer:1993a}. The evolution equation for this trajectory is called a stochastic master equation (SME) \cite{carmichael:1993a,gardiner:2004b,wiseman:2001a,molmer:1993a,gambetta:2005a}.

In this paper, we consider measurement in circuit quantum electrodynamics (QED)~\cite{blais:2004a,wallraff:2004a,wallraff:2005a,schuster:2005a,gambetta:2006a}. This system consists of a cooper pair box, playing the role of an artificial atom, dispersively coupled to a 1-D transmission line resonator and is the circuit equivalent of cavity QED.  It has the advantage that the qubit can be fixed at an antinode of the resonator which, due to its 1D configuration, has a very large vacuum electric field.  This leads to very strong Jaynes-Cummings type coupling between the qubit and the resonator.  This allows to probe a new regime of parameter space, where the resonator and qubit are {\em strongly} coupled via a dispersive interaction and no energy is exchanged between them.  In this regime, the qubit causes a large state-dependent shift of the resonator frequency and thus by monitoring the signal transmitted through the resonator we can infer the qubit's state.  An important feature of the quasi-one-dimensional circuit resonator is that it permits dispersive couplings $\sim 10^6$ times larger than for ordinary three-dimensional cavities.  Coupling of superconducting charge qubits to 3D microwave cavities has also been investigated theoretically~\cite{you:2003a}.

To go beyond previous theoretical works on this system, we employ quantum trajectory theory~\cite{carmichael:1993a,gardiner:2004b,wiseman:2001a,molmer:1993a,gambetta:2005a} and a polaron transformation~\cite{mahan:2000a} to derive an effective SME for the qubit.  It is well know in quantum optics that if the voltage of a cavity is weakly monitored, then the evolution of the conditional state of the cavity is given by the homodyne SME~\cite{carmichael:1993a,wiseman:1993a}. Using this SME as a starting point, we show that if the rate at which information is coming out of the resonator is much larger then the rate at which information is being lost into unmonitored baths, then we can use a polaron transformation to eliminate the resonator and obtain a SME for the qubit only.

The effective SME which is derived corresponds to a weak dispersive measurement of the qubit observable $\op \sigma_z$. This is akin to SMEs derived in Refs.~\cite{korotkov:1999b,korotkov:2001a,korotkov:2001b,wiseman:2001b,goan:2001a,korotkov:2003a,oxtoby:2005a,oxtoby:2007a} for a quantum dot monitored by a quantum point contact.  In Refs.~\cite{korotkov:1999b,korotkov:2001a,korotkov:2001b,korotkov:2003a}, the SMEs are referred to as the quantum Bayesian equations.  They can be seen as the Stratonovich version of the the \ito SMEs presented here and in Refs.~\cite{goan:2001a,oxtoby:2007a}.  We note that one can derive a similar equation by adiabatically eliminating the resonator from the qubit-resonator master equation using a similar technique as presented in \cite{doherty:1999a}.  However, by using the present method, we are able to derive corrections to the various system rates.  These rates are shown to agree very well with the solution of the total resonator-qubit conditional state found by numerically solving the homodyne SME.

The paper is organized as follows. The next section contains a brief discussion of the circuit QED hamiltonian and of the dispersive approximation.  In Sec.~\ref{sec:MasterEq}, we use a polaron transformation to eliminate the resonator from the resonator-qubit master equation and in this way obtain an exact master equation for the qubit only.  This exact effective master equation shows how the resonator induces an additional dephasing channel on the qubit, whose strength depends on the amplitude of input drives on the resonator.  This dephasing rate corresponds to measurement-induced dephasing found in Ref.~\cite{gambetta:2006a}.  It contains the number splitting predicted in Ref.~\cite{dykman:1987a,gambetta:2006a} and experimentally reported in Ref.~\cite{schuster:2007a}.  In Sec.~\ref{sec:QT} we apply the same polaron transformation to derive an effective SME for the qubit.  This equation corresponds to a weak measurement of the qubit $\op\sigma_z$ operator with rate $\Gamma_\mathrm{ci}$ and extra {\em non-Heisenberg} backaction  which causes random rotations of the qubit around the $\op\sigma_z$ with rate $\Gamma_\mathrm{ba}$.  Using these results, in Sec.~\ref{sec:meas} we investigate the situation where the measurement is quantum limited and how a measurement time can be defined.  In Sec.~\ref{sec:jumps}, we describe the emergence of quantum jumps in the conditional state as the measurement strength is increased.  In particular, we investigate how the spontaneous relaxation into the qubit bath reveals itself as a jump when the measurement is strong, rather than a diffusive decay when the measurement is weak.
Finally, in Sec.~\ref{sec:zeno}, we include a qubit control drive in our description of the system and investigate how the Zeno effect \cite{breuer:2002a,Gagen:1993a,Presilla:1996a,Cresser:2006a} can be observed in this system.  We summarize our conclusions in Sec.~\ref{sec:con}.

\section{Cavity QED with superconducting circuits}
\label{sec:circuit_QED}

\begin{figure}[t]
	\centering
		\includegraphics[width=.45\textwidth]{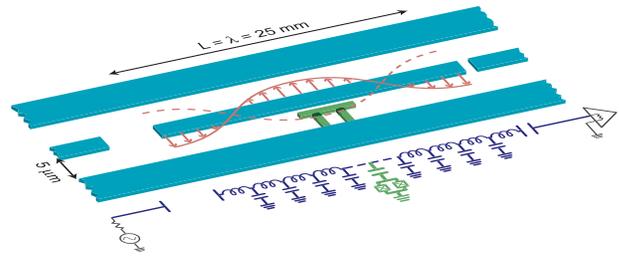}
	\caption{(Color online) Schematic layout and lumped element version of the circuit QED implementation. A superconducting charge qubit (green) is fabricated inside a superconducting 1D transmission line resonator (blue).  Here, we have taken the input (LHS) and output (RHS) capacitance of the resonator to be unequal.  With the output capacitance much larger than the input one, a larger fraction of the photons in the resonator will escape on the output side.  In this way, the signal to be measured is enhanced.}
	\label{fig:CQEDQuantumTrajectoryFig01}
\end{figure}

We consider a Cooper pair box capacitively coupled to a transmission line resonator acting as a simple harmonic oscillator.  This system, illustrated schematically in Fig.~\ref{fig:CQEDQuantumTrajectoryFig01}, was first introduced in Ref.~\cite{blais:2004a} and experimentally studied in Refs.~\cite{wallraff:2004a,schuster:2005a,wallraff:2005a,schuster:2007a,houck:2007a,majer:2007a}.  Measurement-induced dephasing was theoretically studied in detail in Ref.~\cite{gambetta:2006a} and the applications to quantum information processing investigated in Ref.~\cite{blais:2007a}.

As described in the above references, the system's Hamiltonian in the presence of a microwave drive of amplitude $\veps_d(t)$ and frequency $\w{d}$ can be written as~\cite{blais:2004a}
\begin{equation}\label{eq:FreeHam1}
	\begin{split}
 		\op{H}  = &\frac{ \hbar\omega_a}{2}\op{\sigma}_z+ \hbar\omega_r
 		\op{a}\dg\op{a}+  \hbar g \left(\op{a}\dg\op\sigma_- + \op{a}\op{\sigma_+}\right)\\ &
 		 + \hbar \left[\veps_{d}(t)\op{a}\dg e^{-i\omega_d t}+\veps_{d}^*(t)\op{a} e^{i\omega_d t}\right].
	\end{split}
\end{equation}
In this expression, $\wrr$ is the resonator frequency, $\wa$ the qubit transition frequency and $g$ the resonator--qubit coupling strength.  Depending on its frequency, the drive can correspond either to  measurement of the qubit or can be used to coherently control its state.  In the dispersive regime, when $|\Delta| = |\wa-\wrr| \gg |g|$, the effective Hamiltonian~\eq{eq:FreeHam1} can be approximated by \cite{blais:2004a}
\begin{equation}\label{eq:FreeHam3}
 	\op{H}_{\rm eff} =\frac{ \hbar\tilde\omega_a}{2}\op{\sigma}_z+ \hbar\Delta_r
 	\op{a}\dg\op{a}+ \hbar\chi\op{a}\dg\op{a}\op{\sigma}_z+
	 \hbar\left[\veps_{d}(t)\op{a}\dg+\veps_{d}^*(t)\op{a}\right],
\end{equation} where $\Delta_r = \omega_r - \omega_d$ and 
 we have assumed that the drive is far off resonance from the qubit transition frequency and moved the resonator to a frame rotating at frequency $\omega_d$.  In this expression, we have defined $\chi = g^2/\Delta$ as the dispersive coupling strength between the resonator photon number and the qubit.  Moreover, we have taken  $\tilde\omega_a = \wa + \chi$ as the Lamb shifted qubit transition frequency and defined $\Drr = \wrr-\w{d}$.

It is important to note that the dispersive approximation breaks down as the number of photons in the resonator approaches the critical photon number $n_{\rm crit} = \Delta^2/4g^2$~\cite{blais:2004a}.   In the present paper, we will work at moderately low photon number and assume the dispersive approximation to hold. That is, we will assume that Eq.~\eqref{eq:FreeHam3} is a valid description of the system.  The results obtained will therefore also apply to the circuit QED implementation that uses the so-called transmon qubit~\cite{koch:2007a}.  A future publication will explore the break down of the dispersive approximation~\cite{boissonneault:2007a} at small detuning and large photon number.

\section{Master Equation}\label{sec:MasterEq}

\subsection{Master equation for the combined system}

In the Born-Markov approximation, the master equation describing circuit QED takes the usual Lindblad  form \cite{lindblad:1976a,carmichael:1993a}
\begin{equation}\label{eq:MasterEq}
	\begin{split}
 		\dot{\crho}(t)=& -\frac{i}{\hbar}\left[\op{H}_{\rm eff},\crho(t)\right]+\kappa{\cal D}[\op{a}]\crho(t)+\gamma_1{\cal D}[\op{\sigma}]\crho(t)\\&
 		+\gamma_{\phi}{\cal  D}[\op{\sigma}_z]\crho(t)/2\\
 		=&{\cal L}_{\rm tot} \crho(t),
	\end{split}
\end{equation}
where $\crho(t)$ is the state matrix for both the qubit and the resonator and ${\cal D}[\op{A}]$ is the damping superoperator defined by the mapping
\begin{equation} \label{eq:D}
   {\cal D}[\op{A}]\crho= \op{A}\crho\op{A}\dg-\op{A}\dg\op{A}\crho/2-\crho\op{A}\dg\op{A}/2.
\end{equation}
The three damping channels are photon loss through the resonator ($\kappa$), qubit decay ($\gamma_1$) and dephasing of the qubit ($\gamma_{\phi}$) \footnote{In the dispersive regime, the operators describing damping, dephasing and measurement should by transformed in the same way as the Hamiltonian. This leads to corrections at order ($g/\Delta$) and can be ignored here. These effects where briefly described in \cite{blais:2004a} and are investigated in detail in~\cite{boissonneault:2007a}.}.

There are two distinct contributions to the resonator damping: loss of photon from the input port described by the rate $\kappa_\mathrm{in}$ and loss of photons at the output port $\kappa_\mathrm{out}$.  The sum of these two rates  $\kappa =\kappa_\mathrm{in} + \kappa_\mathrm{out}$ is what appears in \eq{eq:MasterEq}.  The advantage of distinguishing these two contributions is that, in current experiments, only the photons leaking out of the output port are monitored.  Hence, as we will see later, $\kappa_\mathrm{out}$ is related to the rate at which we acquire information about the qubit state.

\subsection{Master equation for the qubit}

In Ref. \cite{gambetta:2006a} the solution of the above master equation, neglecting energy loss due to
$\gamma_1$ (but keeping its effect on dephasing), was found to be
\begin{equation}
	\crho(t) = \sum_{i,j = e,g} c_{i,j}(t) \ket{i}\bra{j}\otimes\ket{\alpha_i(t)}\bra{\alpha_j(t)},
\end{equation}
where the indices $g$ and $e$ label the qubit ground and excited states, respectively.  In this expression, the coefficients $c_{i,j}(t)$ are given by
\begin{equation}
	\begin{split}
		c_{e,e}(t) &= c_{e,e}(0)\\
		c_{g,g}(t) &= c_{g,g}(0)\\
		c_{e,g}(t) &= \frac{c_{e,g}(0) e^{-i (\tilde\omega_a - i \gamma_2)t}e^{-i 2 \chi \int_0^t\alpha_e(s)\alpha_g^*(s) ds}}{\langle \alpha_g(t)\ket{\alpha_e(t)}}\\
		c_{g,e}(t) &= c_{e,g}^*(t)
	\end{split}
\end{equation}
and $\ket{\alpha_{e(g)}(t)}$ are coherent states of the resonator with amplitudes determined by
\begin{equation}\label{eq:alpha}
	\begin{split}
		\dot\alpha_e(t)&= -i\veps_d(t)-i(\Delta_r+\chi)\alpha_e(t)-\kappa\alpha_e(t)/2 \\
		\dot\alpha_g(t)&= -i\veps_d(t)-i(\Delta_r-\chi)\alpha_g(t)-\kappa\alpha_g(t)/2.
	\end{split}
\end{equation}
In the expression for $c_{e,g}(t)$, we see that in addition to decay due to dephasing $\gamma_2 = \gamma_1/2+\gamma_\phi$, the off-diagonal element of the qubit density matrix decays at a rate that depends on the field amplitudes $\alpha_g$ and $\alpha_e$.  Since the coherent states with these amplitudes act as pointer states in the measurement of the qubit, this decay can be interpreted as measurement-induced dephasing and will depend on how distinguishable the states $\ket{\alpha_e(t)}$ and $\ket{\alpha_g(t)}$ are~\cite{gambetta:2006a}.  Following Ref.~\cite{gambetta:2006a}, the effect of qubit relaxation in this model has been described by Bonzom {\it et al.}~\cite{bonzom:2007a}.

\begin{figure}[tbp]
	\centering
		\includegraphics[width=0.45\textwidth]{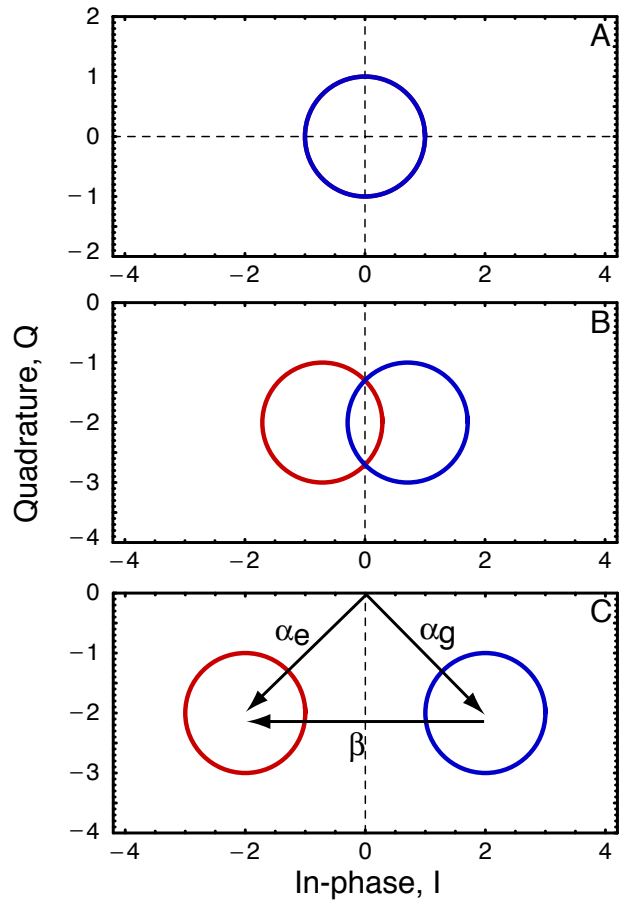}
	\caption{(Color online) In-phase $I = \mathrm{Re}[\langle\op a\rangle]= \langle \op a +\op a\dg \rangle/2$ and quadrature $Q = \mathrm{Im}[\langle \op a\rangle] = \langle i\op a\dg -i\op a\rangle/2$ component of the Q-function solution for the resonator state $\mathrm{Tr}_{\mathrm{qubit}}[\crho_{s}]$.  Here the parameters are $\kappa/2\pi = 10$ MHz, $\chi/2\pi = 5$ MHz, $\epsilon/2\pi = 20$ MHz, and $\Delta_r = 0$.  The measurement drive is turned on at $t=$50 ns and the initial state is $\ket{0}\otimes (\ket{e} + \ket{g})/\sqrt{2}$.  For illustration purposes, $\gamma_1$ was taken to be zero.  A) $t=0$, B) $t=75$ ns, C) Steady-state solution.}
\label{fig:CQEDQuantumTrajectoryFig02}
\end{figure}

How these coherent states act as pointer states for the qubit is illustrated in Fig.~\ref{fig:CQEDQuantumTrajectoryFig02}, where a phase-space representation of the resonator state $\mathrm{Tr}_{\mathrm{qubit}}[\crho(t)]$ is plotted for various qubit--field interaction times.  In these plots, the initial state is taken to be the vacuum for the resonator and the symmetric superposition $(\ket{e} + \ket{g})/\sqrt{2}$ for the qubit.   A continuous coherent drive is turned on at time $t=50$~ns to build up photon population of the resonator.   The three panels of Fig.~\ref{fig:CQEDQuantumTrajectoryFig02} illustrate how the two coherent states $\ket{\alpha_e(t)}$ and $\ket{\alpha_g(t)}$ eventually become separated in phase-space.  Homodyne detection of the resonator field, with the proper choice of local oscillator phase $\phi$, can then be used to distinguish between these two coherent states and thus readout the state of the qubit.   In the last panel of Fig.~\ref{fig:CQEDQuantumTrajectoryFig02} we have introduced the distance between the states $\ket{\alpha_e(t)}$ and $\ket{\alpha_g(t)}$
\begin{equation}
	\beta(t) = \alpha_e(t) - \alpha_g(t).
\end{equation}
Another useful quantity is the angle
\begin{equation}\label{eq:theta}
	\theta_\beta = \arg(\beta).
\end{equation}
For example, for the case of Fig.~\ref{fig:CQEDQuantumTrajectoryFig02}C), this angle is $\theta_\beta = \pi$, which corresponds to the quadrature containing the most information about the state of the qubit.  Indeed, for this particular case, all of the information is in the in-phase component $I$, with no information stored in the quadrature $Q$ component.

Our goal in the remainder of this section is to obtain an effective master equation for the qubit by eliminating the resonator degree of freedom from the full master equation,~\eq{eq:MasterEq}.  To achieve this, we first go to a frame defined by the transformation
\begin{equation} \label{Uframe}
	\tP(t)= \op\Pi_e\op{D}[\alpha_e(t)]+\op\Pi_g\op{D}[\alpha_g(t)]
\end{equation}
with $\op{D}[\alpha]$ the displacement operator of the resonator,
\begin{equation} \label{displace}
	\op{D}[\alpha]=\exp[\alpha \op{a}\dg-\alpha^*\op{a}]
\end{equation}
and $\op\Pi_{j} = \ket{j}\bra{j}$ projectors on the ground and excited states of the qubit.  This is similar to the polaron transformation which has been used extensively in various systems~\cite{mahan:2000a,leggett:1987a}.  For example, this was used in Ref. \cite{irish:2005a} to study a charge qubit coupled to a mechanical oscillator beyond the rotating wave approximation.  However, here we use the transformation~\eq{Uframe} on the dispersive Hamiltonian, not on the full Jaynes-Cummings Hamiltonian.

As is shown in Appendix~\ref{frame}, applying this transformation on the master equation \eq{eq:MasterEq} and tracing over the resonator state yields the laboratory frame reduced qubit master equation
\begin{equation}\label{eq:MasterEqQubit}
	\begin{split}
		\dot{\qrho}(t)=& -i\frac{\omega_{\rm ac}(t)}{2}\left[\sz, \qrho(t)\right]
    	+\gamma_1{\cal D}[\smm] \qrho(t)
    	\\&+\left[\gamma_{\phi} + \Gamma_\mathrm{d}(t)\right]{\cal D} [\sz] \qrho(t)/2\\
    	=&{\cal L} \qrho(t),
	\end{split}
\end{equation}
where $\qrho(t) = {\rm Tr}_{\rm res}[\crho(t)]$.  This expression is the main result of this section.  It is important to note that, within the dispersive approximation, this result is exact.

In this reduced description, we see that the effect of coupling to the resonator translates into an additional dephasing rate $\Gamma_\mathrm{d}(t)$ given by
\begin{equation}\label{eq:Gamma_t}
	\Gamma_\mathrm{d} (t) = 2\chi {\rm Im}[\alpha_g(t)\alpha_e^*(t)].
\end{equation}
In addition to dephasing, the qubit transition frequency is also modified by the photon population of the resonator.  The shifted qubit frequency is given by
\begin{equation}
	\wac = \tilde{\omega}_a + B(t)
\end{equation}
with
\begin{equation}
	B(t)=2\chi\mathrm{Re}[\alpha_g(t)\alpha_e^*(t)].
\end{equation}
This last term gives rise to the ac-stark shift experimentally measured in Ref.~\cite{schuster:2005a}.  In the situation where $\chi\gg \kappa$, the above leads to the number splitting predicted in Ref.~\cite{dykman:1987a,gambetta:2006a} and experimentally observed in Ref.~\cite{schuster:2007a}.

\begin{figure}[t]
	\centering
		\includegraphics[height=3in]{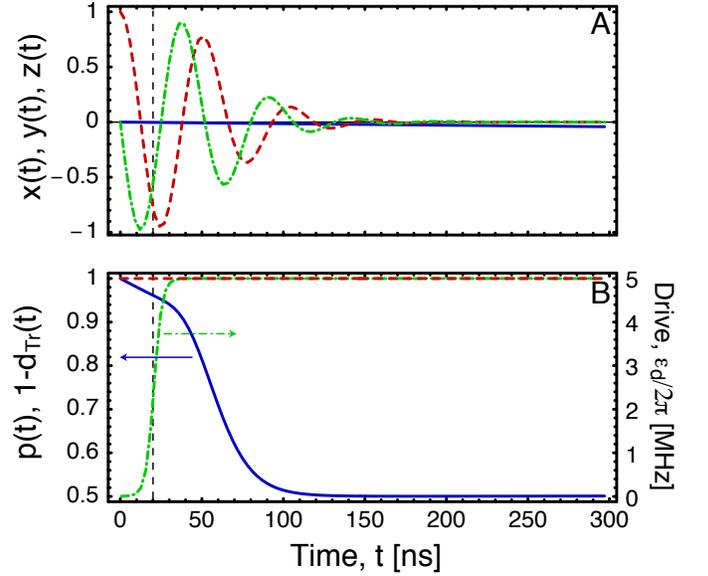}
	\caption{(Color online) A) Numerical solution of the three components of the Bloch vector as obtained from the full master equation: $x$ red dashed, $y$ green dashed-dotted and $z$ solid blue.  For this simulation, a measurement drive tuned at the resonator frequency ($\Delta_r =0$) and with envelope $\varepsilon_\mathrm{d}(t) = A\tanh[(t - t_\mathrm{on})/\sigma]$ is turned on at $t_\mathrm{on}=20$ ns (green dashed-dotted line in panel B).  The turn-on time is indicated with the vertical dashed line.  B) Purity $p$ (solid blue line) and the trace distance $d_\mathrm{Tr}$ (red dashed line) with respect to the effective model derived here.  The latter shows the excellent agreement between the model and full numerical results.  For these results, we have taken $A /2\pi=5$ MHz, $\sigma =5$ ns,  $T_1 = 7\:\mu$s and $T_2=500$ ns.  The other parameters are the same as in Fig. \ref{fig:CQEDQuantumTrajectoryFig02}.}
	\label{fig:CQEDQuantumTrajectoryFig03}
\end{figure}

Although the above results are analytically exact, we have compared the numerical solution of the qubit's dynamics obtained from the full master equation~\eq{eq:MasterEq} to that obtained from the reduced model~\eq{eq:MasterEqQubit}.   This is shown in Fig.~\ref{fig:CQEDQuantumTrajectoryFig03} where the elements of the Bloch vector are plotted as a function of time.  It is useful to note that,  in terms of the Bloch vector
\begin{equation}
	\qrho(t)=\frac{1}{2}\left(\op{1} + x(t)\op\sigma_x+y(t)\op\sigma_y+z(t)\op\sigma_z\right),
\end{equation}
where $x(t)=\mathrm{Tr}[\op\sigma_x\qrho(t)]$, $y(t)=\mathrm{Tr}[\op\sigma_y\qrho(t)]$, and $z(t)=\mathrm{Tr}[\op\sigma_z\qrho(t)]$, the effective master equation,~\eq{eq:MasterEqQubit}, reduces to the simple Bloch equations
\begin{equation}\label{eq:MasterEqQubitblock1}
	\begin{split}
		\dot{x}(t)&=-\wac y(t)-[\gamma_2+\Gamma_\mathrm{d}(t)] x(t)\\
		\dot{y}(t)&=\wac x(t)-[\gamma_2+\Gamma_\mathrm{d}(t)] y(t)\\
		\dot{z}(t)&=-\gamma_1(z(t)+1).
	\end{split}
\end{equation}
Not surprisingly, as seen from Fig.~\ref{fig:CQEDQuantumTrajectoryFig03}, the agreement between numerical integration of these Bloch equations and full numerical integration of the full master equation is excellent.  The agreement between these two results can be quantified by calculating the trace distance $d_\mathrm{Tr}(t)$ between the full and the effective model.  The trace distance in this case is  defined as~\cite{nielsen:2000a}
\begin{equation}
d_\mathrm{Tr}(t) = \frac{1}{2} \mathrm{Tr} \left[\left| \qrho(t)-\mathrm{Tr}_{\mathrm{res}}[\crho(t)] \right|\right],
\end{equation}
and is found to be zero at all integration times, up to numerical round off (truncation) errors. This was checked for a wide range of measurement amplitudes $\veps_{d}$ and the trace distance was also found to be zero for all verified amplitudes.  Also in this figure is shown the purity (full blue line)
\begin{equation}
p(t) = \mathrm{Tr}[\qrho(t)^2] = \frac{1}{2}[1 +x(t)^2 +y(t)^2 +z(t)^2].
\end{equation}
{ In this master equation description, the purity tends to 1/2, corresponding to a completely mixed state, due to dephasing.   On top of this dynamics, relaxation is taking the purity to 1 (a pure state) at a rate $\gamma_1$.  For the parameters chosen in Fig.~\ref{fig:CQEDQuantumTrajectoryFig03} relaxation is much weaker than (measurement-induced) dephasing, such that only the effect of the latter is seen.}

To summarise this section, we have obtained an {\em exact} master equation description of the qubit dynamics that only involves classical solution of the resonator field.   In addition to giving direct insights in the qubits dynamics (e.g. measurement induced dephasing and ac-Stark shift) this also provides a tool to significantly reduce the complexity of numerical calculations.  Indeed, only a 2-dimensional Hilbert space is now required.    We note that this effective model was used to analyze and reproduce, with exceptional agreement, the experimental results reported in Ref.~\cite{houck:2007a}.

\section{Stochastic Master equation}\label{sec:QT}

\subsection{Stochastic master equation for the combined system}

In this section, we review homodyne measurement of the field emitted from the resonator.  Using these results, we will in the next subsection use the transformation of \eq{Uframe} to obtain an effective SME for the qubit only.

In a given quantum system, if all the decoherence and decay channels can be monitored continuously, it is possible to describe the system conditioned on the results  of this monitoring result  $J(t)$ by a pure state $\ket{\Psi_J}$ rather than by the average state $\crho$. This pure states is called a conditional state and can be viewed as our state of knowledge of the system. However, in systems where information about only some of the decay channels can be obtained, there is missing information, and it is no longer possible to have a pure state description.  In this case, we can assign a conditional state matrix, $\crho_J$ to represent the state of the system under continuous observation of the particular decay channels.
The evolution equation of this conditional state is referred to as a stochastic master equation (SME)  \cite{carmichael:1993a, molmer:1993a, wiseman:1993a, gardiner:2004b, wiseman:2001a, korotkov:2001a, korotkov:2003a, gambetta:2005a}.  It has the property that its average state $\crho(t)$,
\begin{equation}
	\crho(t) = \mathrm{E}[\crho_J(t)]
\end{equation}
is the solution to the master equation. Here $\mathrm{E}$ denotes an ensemble average over measurement records $J(t)$

In circuit QED, there are four decay channels described by the rates $\kappa_\mathrm{in}$, $\kappa_\mathrm{out}$, $\gamma_1$ and $\gamma_\phi$.  Only information coming out of the $\kappa_\mathrm{out}$ channel, that is photons transmitted through the resonator, are monitored.  As a result, we must restrict our description to a conditional density matrix.   As will be discussed in the next section, in the effective model for the qubit only, this will correspond to monitoring the decay channel corresponding to the rate $\Gamma_\mathrm{d}(t)$ and not monitoring the $\gamma_1$ and $\gamma_\phi$ channels.

Although,  direct detection of the transmitted microwave photons is possible~\cite{houck:2007a}, here we will consider homodyne processing.  That is, we will assume that the signal coming from the output port of the resonator is mixed with a strong local oscillator of phase $\phi$ tuned to the signal frequency.  Given the homodyne measurement result $J(t)$, we can assign to the qubit--resonator system the conditional state $\crho_J(t)$ whose evolution is governed by the SME~\footnote{In Ref. \cite{wiseman:1993a} only Lindblad superoperators with one decoherence term are considered and as such a pure state unraveling is possible,  here however we have three decoherence terms and only unravel one of them. That is, we have a partial measurement and must use a SMEs unless we use the numerical techniques of Ref. \cite{gambetta:2005a}.}
\begin{equation}
	\begin{split}\label{eq:HomodyneQT}
		\dot\crho_J(t)
		=&{\cal L}_{\rm tot}\crho_J(t)
		+i\sqrt{\kappa\eta}[\op Q_\phi,\crho_J](J(t) -\sqrt{\kappa\eta}\langle2\op I_\phi\rangle_t)\\&+\sqrt{\kappa\eta}\Mso{2\op I_\phi}{o}\crho_J(t)[J(t) -\sqrt{\kappa\eta}\langle2\op I_\phi\rangle_t]\\
	\end{split}
\end{equation}
with ${\cal L}_{\rm tot}$ given by \eq{eq:MasterEq}, $\eta$ is the measurement efficiency which we define below, $\Mso{\op c}{\langle \op c\rangle}$ is the measurement superoperator defined as
 \begin{equation}
	\Mso{\op{c}}{\langle \op c\rangle}\crho= (\op{c} -\langle\op{c}\rangle_t) \crho/2+\crho(\op{c}-\langle\op{c}\rangle_t)/2,
\end{equation}
where $\langle\op{c}\rangle_t = \mathrm{Tr}[\op c \crho_J(t)]$ and
the $\phi$-dependent field components are $2\op I_\phi = \op a e^{-i\phi} + \op a \dg e^{i\phi}$ and $2\op Q_\phi = -i\op a e^{-i\phi} + i\op a \dg e^{i\phi}$. The efficiency is $\eta =\kappa_\mathrm{out}\eta_\mathrm{det}/\kappa$ with $\kappa_\mathrm{out}$ the rate at which photons come out of the output port of the resonator and $\eta_\mathrm{det}$ is the  efficiency at which these photons are detected.  For the current circuit QED experiments~\cite{wallraff:2004a,schuster:2005a,wallraff:2005a,schuster:2007a,houck:2007a,majer:2007a}, this can be written as $\eta_\mathrm{det} = 1/(N+1)$ with $N$ the number of noise photons added by the amplifier stage.

For homodyne detection, the measurement record observed in an experiment can be expressed as
\begin{equation}
	J(t) =\sqrt{\kappa\eta} \langle2\op I_\phi \rangle_t +\xi(t),
\end{equation}
where $\xi(t)$ is Gaussian white noise and represents the photon shot noise. It is formally defined as~\cite{gardiner:1985a}
\begin{eqnarray}
	{\rm E}[\xi(t)]&=&0\\
	{\rm E}[\xi(t)\xi(t')]&=&\delta(t-t')
\end{eqnarray}
with ${\rm E}$ denoting an ensemble average over realizations of the noise $\xi(t)$.

The measurement term in \eq{eq:HomodyneQT}, the one including the superoperator $\Mso{2\op I_\phi}{\langle \op c\rangle}$, comprises two parts.  We will refer to the first one as the homodyne gain and the second as the innovation. The homodyne gain is $\sqrt{\kappa\eta}\Mso{2\op I_\phi}{\langle\op I_\phi\rangle}\crho_J(t)$ and collapses the state towards a $I_\phi$ state (eigenstate of $I_\phi$). The innovation is $J -\sqrt{\kappa\eta}\langle 2\op I_\phi \rangle_t$, it pushes the conditional state to a higher or lower $I_\phi$ state depending on whether the current result $J(t)$ is greater or smaller then the average $\sqrt{\kappa\eta}\langle 2\op I_\phi\rangle_t$.
The last term in \eq{eq:HomodyneQT}, the noisy Hamiltonian term is extra non-Heisenberg backaction (it does not come with any information gain) caused by the measurement. It is this term that stops the conditional state from being driven to a $I_\phi$ state as it supplies random  $Q_\phi$ kicks to the conditional state causing delocalization in $I_\phi$.

For heterodyne measurement (where the local oscillator (LO) is detuned from the signal frequency), both component $\op I_\phi$ and $\op Q_\phi$ can be measured simultaneously.  In this case, the SME \eq{eq:HomodyneQT} has extra gain and innovation terms corresponding to this additional component.  Heterodyne measurement also lowers the measurement efficiency $\eta_\mathrm{det}$ by a factor of $1/2$.  Detection of this second component does not change the physics in an essential way, and we will focus on the simpler case of homodyne detection in the remainder of this paper.

Finally, we note that the SME \eq{eq:HomodyneQT} is known to lead to a diffusive-like evolution for the system~\cite{wiseman:2001a,wiseman:1993a,wiseman:1993b}.  We will however see in section~\ref{sec:jumps} how there can be a cross-over from diffusive to jump-like evolution as the measurement strength is increased.

\subsection{Stochastic master equation for the qubit}

In this section, we use the transformation \eq{Uframe} to obtain an effective SME for the qubit only.  As described in Appendix \ref{app:qt}, this can be done in the limit where
\begin{equation} \label{eq:coherent_info_limit}
	\epsilon = \frac{2\gamma_1}{\kappa}\ll 1.
\end{equation}
We note that this limit does not imply that the information gain about the state of the qubit is small.   For the system to be in this limit, we simply require that the photons come out of the resonator faster than the qubit decay rate. Otherwise, the full SME of \eq{eq:HomodyneQT} must be used.  However, as will be seen from numerical investigations, the validity of the model can extend well beyond this limit in practice.

As described in Appendix \ref{app:qt}, in the limit where \eq{eq:coherent_info_limit} is valid, the effective SME for the qubit obtained using the transformation of \eq{Uframe} and tracing over the resonator states is
\begin{equation}\label{eq:qubitQT}
\begin{split}
\dot{\qrho}_\Jp(t)=&{\cal L} \qrho_\Jp(t)+\sqrt{\Gamma_\mathrm{ci}(t)}\Mso{\szo}{\langle\szo\rangle}\qrho_\Jp(t) (\Jp(t)
- \sqrt{\Gamma_\mathrm{ci}(t)}\langle \op\sigma_z \rangle_t)
\\
& -i \frac{\sqrt{\Gamma_\mathrm{ba}(t)}}{2} [\szo,\qrho_\Jp(t)] (\Jp(t) - \sqrt{\Gamma_\mathrm{ci}(t)}\langle \op\sigma_z \rangle_t).
\end{split}
\end{equation}
In this expression, $\Gamma_\mathrm{ci}(t)$ is the rate at which coherent information comes out the resonator and $\Gamma_\mathrm{ba}(t)$
represents extra non-Heisenberg back-action from the measurement.  These rates are given by
\begin{align}
\Gamma_\mathrm{ci}(t) =& \eta\kappa|\beta(t)|^2\cos^2(\phi - \theta_\beta ) \label{eq:GammaCI}\\
\Gamma_\mathrm{ba}(t) =&\eta\kappa|\beta(t)|^2\sin^2(\phi-\theta_\beta),\label{eq:GammaBA}
\end{align}
with the angle $\theta_\beta$ being defined in \eq{eq:theta}.

In \erf{eq:qubitQT}, $\Jp(t)$ is the processed record coming from the resonator and is given by
\begin{equation}\label{eq:normcurrent}
\Jp(t) =  \sqrt{\Gamma_\mathrm{ci}}\langle \op\sigma_z \rangle_t +  \xi(t).
\end{equation}
which can be related to the homodyne current by
\begin{equation}\label{eq:relation}
{J}(t) =   \Jp (t)+ \sqrt{\kappa\eta}|\mu(t)| \cos(\theta_\mu-\phi)
\end{equation}
with $\mu(t) = \alpha_e(t) + \alpha_g(t)$ and $\theta_\mu=\arg(\mu)$.

From the above expressions, by doing homodyne measurement on the field leaving the resonator we are in fact doing a weak diffusive measurement of the qubit operator $\op{\sigma}_z$ with measurement strength $\Gamma_\mathrm{ci}(t)$, as well as  causing extra unitary back-action on the qubit (by the term proportional to $\Gamma_\mathrm{ba}(t)$). This is clearly seen by rewriting \erf{eq:qubitQT} in terms of the Bloch vectors
\begin{equation}
\label{eq:blockQTz}
\begin{split}
	\dot{x}_J(t)=&-\{\omega_{\rm ac}(t) + \sqrt{\Gamma_\mathrm{ba}(t)}[\Jp(t)- \sqrt{\Gamma_\mathrm{ci}(t)}z_J(t)]\}y_J(t)\\& -\{\gamma_2+\Gamma_\mathrm{d}(t) +\sqrt{\Gamma_\mathrm{ci}(t)}z_J(t)[\Jp-\sqrt{\Gamma_\mathrm{ci}(t)}z_J]\}  x_J(t) \\
	        \dot{y}_J(t)=&\{\omega_{\rm ac}(t) + \sqrt{\Gamma_\mathrm{ba}(t)}[\Jp(t)- \sqrt{\Gamma_\mathrm{ci}(t)}z_J(t)]\}x_J(t)\\& -\{\gamma_2+\Gamma_\mathrm{d}(t) +\sqrt{\Gamma_\mathrm{ci}(t)}z_J(t)[\Jp-\sqrt{\Gamma_\mathrm{ci}(t)}z_J]\}  y_J(t) \\
	         \dot{z}_J(t)=& \sqrt{\Gamma_\mathrm{ci}(t)} [1-z_J(t)^2] [\Jp(t) - \sqrt{\Gamma_\mathrm{ci}(t)}z_J(t)]\\&-\gamma_1[z_J(t)+1].
\end{split}
\end{equation}
From these expression, we see that the effect of the measurement consists of two parts.  First, a nonlinear update to the state which results in pushing the system towards either $z=\pm1$ depending on the observed record (the terms proportional to $\Gamma_\mathrm{ci}(t)$), and second, an extra stochastic component that causes rotations around the $\op\sigma_z$ axis (the $\Gamma_\mathrm{ba}(t)$ terms). We find that the sum of these rates is $\eta\Gamma_\mathrm{m}(t)$ where
\begin{equation}\label{eq:Gamma_m}
	 \Gamma_\mathrm{m}(t)=\kappa |\beta(t)|^2
\end{equation}
is the maximum measurement rate.
This rate can be simply understood by noting that $\kappa$ is the rate at which photons leak out the resonator and $|\beta(t)|^2$ is the amount of information about the qubit state encoded by these photons. That is, this is the maximum amount of information that can be gained from this system via homodyne monitoring. Looking at Eq.~\eqref{eq:GammaCI} this is achieved by setting the homodyne phase to $\theta_\beta$ and physically corresponds to detecting the quadrature which has the greatest separation of the pointer states (see Fig. \ref{fig:CQEDQuantumTrajectoryFig02}C)). If we set $\phi$ to be $\theta_\beta+\pi/2$, then $\Gamma_\mathrm{ba}(t)=\eta\Gamma_\mathrm{m}(t)$ which results in the measurement revealing no information about the qubit state and only causing dephasing by qubit frequency change via photon shot noise.

\begin{figure}[t]
	\centering
		\includegraphics[width = 0.45\textwidth]{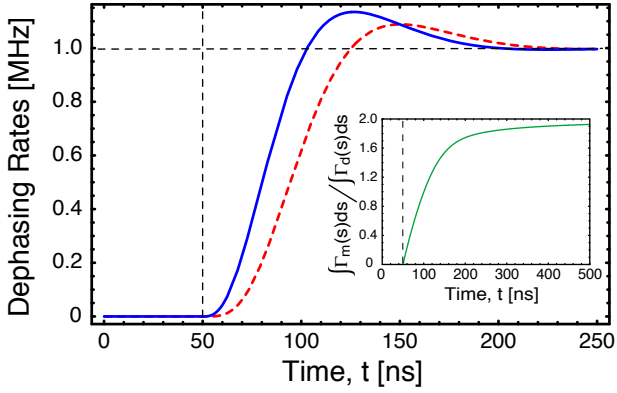}
	\caption{(Color online) $\Gamma_\mathrm{d}(t)$ (solid blue line) and  $\Gamma_\mathrm{m}(t)/2$ (dashed red line) for a measurement drive turned on at the time $t=50$ ns.  The drive frequency is tuned to the bare resonator frequency ($\Delta_r =0$) and the amplitude is $\varepsilon_\mathrm{m}/ 2 \pi = \sqrt{5}$. The rest of the parameters are the same as in Fig. \ref{fig:CQEDQuantumTrajectoryFig02}. The inset shows the ratio between the total dephasing and the total measurement-induced dephasing. The vertical black dashed line is the turn on time for the measurement drive.  The horizontal dashed line is the steady-state dephasing rate.}
	\label{fig:CQEDQuantumTrajectoryFig04}
\end{figure}

When the resonator dynamics have reached steady-state, it can be shown that
\begin{equation}
	\Gamma_\mathrm{d}(t\rightarrow\infty) = \Gamma_\mathrm{m}(t\rightarrow\infty)/2.
\end{equation}
This is illustrated in Fig. \ref{fig:CQEDQuantumTrajectoryFig04} where both $\Gamma_\mathrm{d}(t)$ (solid blue line) and  $\Gamma_\mathrm{m}(t)/2$ (dashed red line) are presented for the situation where a  measurement drive tuned to the bare resonator frequency is turned on at $t=50$ ns.  From this figure, we see that after the resonator transients decay away, the dephasing rate $\Gamma_\mathrm{d}$ is twice the maximum measurement rate $\Gamma_\mathrm{m}$. Also shown in the figure (see inset) is the ratio between the total dephasing of the system, which we define as $\int_0^t \Gamma_\mathrm{d}(s)ds$, and the total measurement-induced dephasing, defined as $\int_0^t \Gamma_\mathrm{m}(s)ds/2$.   As it should (when $\kappa\gg \chi$), the former exceeds, or is at least equal to, the latter.  This is simply a statement that there is additional information about the qubit available in the initial transients of the resonator which is not being used when simply monitoring the homodyne signal. 

\begin{figure}[t]
	\centering
		\includegraphics[width=0.45\textwidth]{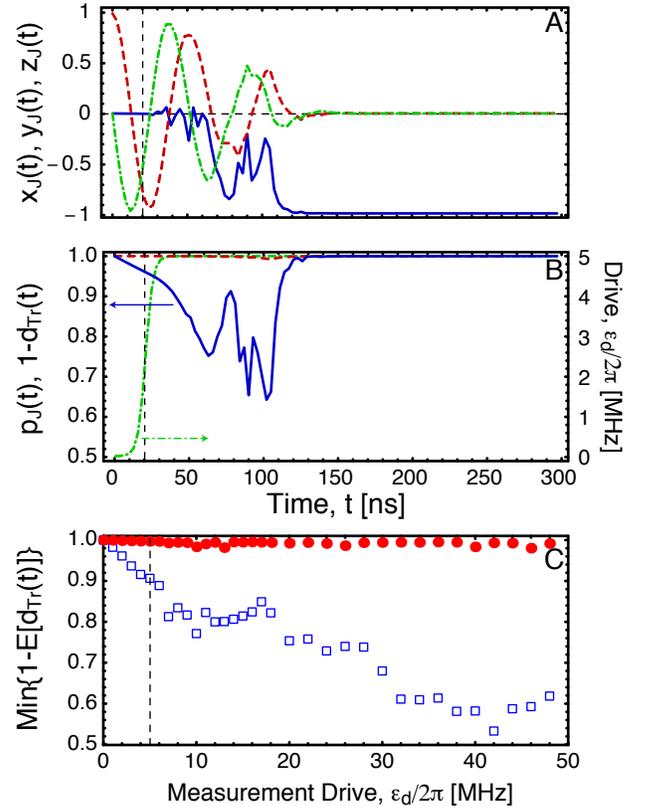}
	\caption{(Color online) A) Numerical solution of the three components of the Block vector as obtained from the full homodyne SME: $x$ red dashed, $y$ green dashed-dotted and $z$ solid blue.  The measurement frequency and envelope is the same as in Fig.~\ref{fig:CQEDQuantumTrajectoryFig03}.   We have assumed perfect efficiency $\eta=1$ and taken the LO phase to be $\phi=\theta_\beta=\pi$, corresponding to maximal information gain at $\Delta_r=0$.  B) Conditional purity $p_J$ (full blue) and measurement pulse amplitude (dot-dashed green).  Also shown is one minus the trace distance (dashed red), as calculated from the full homodyne SME and the effective SME.  C) One minus the trace distance minimized over the simulation time, $300$ ns, for a range of measurement amplitude and two values of $T_1$: 7 $\mu$s (full red dots), 100 ns (empty blues squares).  Here, the trace distance is averaged over 70 trajectories.}
	\label{fig:CQEDQuantumTrajectoryFig05}
\end{figure}

A typical trajectory for the conditional state is shown in Fig. \ref{fig:CQEDQuantumTrajectoryFig05}A)  [red dashed line is the $x$ component, green dashed-doted line is the $y$ component and the solid blue line is the $z$ component] for the case of a measurement drive turned on at $t_\mathrm{on} =20$ ns and of maximal amplitude $\varepsilon_\mathrm{d}^{\mathrm{max}}/2\pi=5$ MHz.  The envelope of the drive, shown as the green dashed-dotted line in Fig.~\ref{fig:CQEDQuantumTrajectoryFig05}B), is the same shape as the one used in Fig. \ref{fig:CQEDQuantumTrajectoryFig03}.

Unlike the average evolution, as we monitor the system we become more certain of its quantum state.  This is represented in Fig.~\ref{fig:CQEDQuantumTrajectoryFig05}A) by the $z$ component being stochastically pushed towards $z=-1$.  Note that we are showing but one possible trajectory, other realizations will tend to localise the $z$ component to +1.  The fact that our state of knowledge about the quantum state is increased by monitoring is also made apparent by the purity which reaches unity [full blue line in panel B)].  Under the average evolution, see Fig.~\ref{fig:CQEDQuantumTrajectoryFig03}b), we end up with a completely mixed state.  This is simply due to the fact that the average description does not take into account the information gain due to the measurement.

Also shown in this panel B) is the trace distance between the conditional state found using the homodyne SME Eq.~\eqref{eq:HomodyneQT} and the effective SME Eq.~\eqref{eq:qubitQT}, in both cases using the same parameters as in panel A).  The agreement between the two results is excellent, which shows that when the limit \eq{eq:coherent_info_limit} is valid, the effective model derived here is indeed a good description.  To see the break down of the validity of the model, we plot in panel C) one minus the trace distance minimized over the simulation time, $300$ ns, as a function of measurement amplitude and for two values of $T_1$.  This is calculated and averaged over 70 trajectories.  For $T_1 = 7\:\mu$s (full red circles), the agreement is very good at all simulated amplitudes.  However, for $T_1 = 100$ ns (empty blue squares), the condition~\eq{eq:coherent_info_limit} is much less valid. As the measurement amplitude is increased, the approximation that we can neglect elements of Eqs. \eqref{eq:b10} and \eqref{eq:b11} in Eq. \eqref{eq:b9} breaks down and the discrepancy between the effective and full models is apparent.  In any cases, it is clear that the effective model is a {\em very} good approximation for long $T_1$ and can give a relatively accurate description of the system even at low $T_1$, provided that the measurement amplitude is not too large.  We note that a $T_1$ of 7 $\mu$s was reported in Ref.~\cite{wallraff:2005a}.  The model obtained here is therefore useful in realistic experimental settings.

Moreover, we find that for the small $T_1$ of 100 ns, while the discrepancy reported in panel C) for an average over many trajectories can be large, when inspecting single trajectories we find no notable new features as the measurement amplitude is increased.  The only change can be attributed to renormalization of the various system parameters under the measurement. This suggests that the effective model presented here may extend well beyond the limit \eqref{eq:coherent_info_limit}, with the only change being that the measurement rate must be obtained numerically (or experimentally measured).

\subsection{Signal-to-noise ratio}

Using the above results, we can define the signal-to-noise ratio (SNR) as 
 \begin{equation}
 {\rm SNR} = \frac{\Gamma_\mathrm{ci}}{\gamma_1} =  \frac{\eta\Gamma_\mathrm{m}\cos^2(\theta_\beta - \phi) }{\gamma_1}.
 \end{equation}
As expected, the SNR is maximized by setting the phase $\phi$ of the local oscillator to the quadrature containing the most information:  $\phi = \theta_\beta$ (see Fig.~\ref{fig:CQEDQuantumTrajectoryFig02}).  Since $\Gamma_\mathrm{m}$ scales with the measurement amplitude, and thus with the number of photons populating the resonator, this expression indicates that arbitrarily large SNRs can be reached in principle by simply increasing this amplitude.  However, as was also pointed out in Ref.~\cite{blais:2004a}, at large photon numbers the dispersive Hamiltonian Eq.~\eqref{eq:FreeHam3} breaks down and the results obtained here are simply no longer valid.

\begin{figure}[t]
	\centering
		\includegraphics[width = 0.45\textwidth]{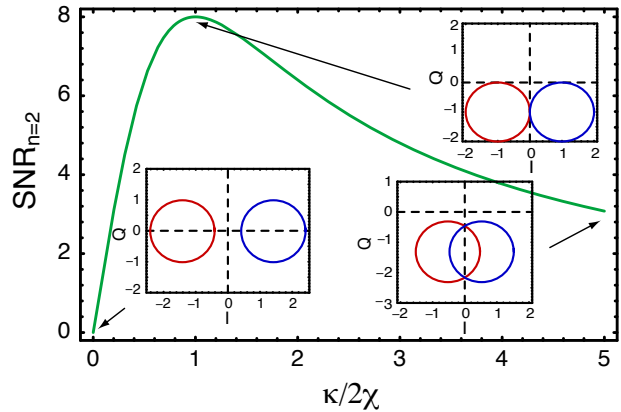}
	\caption{(Color online) The signa-to-noise ratio (SNR) at $\Delta_r = 0$ and for a fixed number of photons in the resonator $n = 2$.  The efficiency was taken to be ideal $\eta = 1$ and the relation rate $ \gamma_1/2\pi = 1$ MHz. The three insets show the phase space representation of the resonator state for $\kappa = 0.01\chi$, $\kappa = 2\chi$ and $\kappa = 5 \chi$, respectively.  The optimal SNR occurs at $\kappa = 2\chi$.}
	\label{fig:CQEDQuantumTrajectoryFig06}
\end{figure}

As a result, we must maximize the SNR at a fixed number of photons which is assumed to be small enough for the dispersive approximation to be valid.  The number of photons inside the resonator depends on the drive amplitude $\varepsilon_d$, on resonator damping $\kappa$, resonator-drive detuning $\Delta_r$ and on the dispersive qubit-resonator coupling $\chi$.  Because of the latter, the actual number of photons can depend on the state of the qubit.  We thus define $n_g$ and $n_e$ the number of photons given that the qubit is in the ground or excited state~\cite{gambetta:2006a}.   When optimizing, we therefore work at a fixed $n= \mathrm{max}(n_e,n_g)$ which is assumed to be well below the dispersive breakdown.  Under this constraint, and assuming that the resonator has reached steady-state, we find that the optimal SNR, for arbitrary $\chi$ and $\kappa$, occurs at $\Delta_r = 0$ and is given by
\begin{equation}
 {\rm SNR} = \frac{ 4 n \eta \kappa\chi^2}{\gamma_1(\kappa^2/4+\chi^2)}.
\end{equation}
A plot of this is shown in Fig. \ref{fig:CQEDQuantumTrajectoryFig06}.  From this plot, it is clear that the optimal value occurs at $\kappa = 2 \chi$ and is given by $\mathrm{SNR_{max}}= 4 n \eta \chi/\gamma_1$.

Thus, to extract the most information about the state of the qubit, we need to tune the measurement drive to bare resonator frequency $\omega_r$ and choose the system parameters such that $\chi = \kappa/2$.   Note that $\kappa$ is chosen at fabrication time (and could be tuned in-situ by a change of resonator design) and that $\chi = g^2/(\omega_a-\omega_r)$ can be tuned in-situ by tuning $\omega_a$.  At a drive strength such that $\bar{n} = n_{\rm crit}$~\cite{blais:2004a}, the maximum SNR is $\eta \Delta /\gamma_1$, which for current experiments is approximately $50 - 100$.  Using the results of Ref.~\cite{gambetta:2007a}, this implies that a measurement fidelity of at least  $95\%$ could be achievable in practice.  To increase further this value, improvements have to be made to $T_1$, to minimise the amplifier noise and/or tune the system to be still further in the dispersive regime.

\section{Measurement time}\label{sec:meas}

In this section, we use the effective SME for the qubit to obtain an estimate for the measurement time.  That is, an estimate for how long it takes us to be confident about the state of the qubit.  To do so, we will take the simplifying assumption that the resonator transient can be be ignored (that is $\Gamma_\mathrm{d} = \Gamma_\mathrm{m}/2$).  In this situation,  
the simplest definition for a measurement time is to assume it is the inverse of the information gain rate:
\begin{equation}\label{eq:measurement_time_simple}
t_\mathrm{m} = 1/\Gamma_\mathrm{ci}.
\end{equation}
Using this definition, we can write
\begin{equation}
t_\mathrm{m}\Gamma_\mathrm{d}\leq 1/2.
\end{equation}
This is the information gain/dephasing `uncertainty relation' already discussed by several authors~\cite{devoret:2000a,averin:2002a,clerk:2003a,makhlin:2001a}.  The equality is saturated for $\eta = 1$ and $\phi = \theta_\beta$ and corresponds to the quantum limit discussed in Ref.~\cite{clerk:2003a}.

In the next sections, we define the measurement time with respect to three different measures.  As we will see however, there is no unique definition of the measurement time.

\subsection {Integrated signal}

For simplicity, let us first discuss the situation where $\gamma_1=0$ and with the qubit initially in either $z=\pm1$.  In this situation, the measurement time is very well defined as the qubit state simply remains in its initial state and the task is to estimate that state in the shortest possible time.

We define the measurement signal for an integration time $t$ as
\begin{equation}\label{eq:signal}
{s}(t)=\sqrt{\Gamma_\mathrm{ci}}\int_0^t  \Jp(t') dt'.
\end{equation}
 This signal has mean and standard deviation
\begin{eqnarray}
\bar{s}(t)&=& \pm t\Gamma_\mathrm{ci}  \\
\Delta |s(t)|&=&\sqrt{\Big{\langle} [ s(t)- \bar{s}(t)]^2
\Big{\rangle}} = \sqrt{{t}{\Gamma_\mathrm{ci}}}.
\end{eqnarray}

As a result,  if we were to measure the system for a time $t$ and repeat this measurement several times,  we would be
confident that, to one standard deviation, the value of ${s}$ is $\pm t\Gamma_\mathrm{ci}\pm \sqrt{t \Gamma_\mathrm{ci}} $. As a result, to be able to distinguish between $z= \pm1$ a reasonable requirement is
\begin{equation} \label{eq:SignalCond}
\Delta |s(t)| \leq  t\Gamma_\mathrm{ci}.
\end{equation}
The equality defines the measurement time to be $t_\mathrm{m} = 1/\Gamma_\mathrm{ci}$.  In this case, we recover the simple assumption discussed above.

A problem with this approach is that the condition which defines distinguishability, and thus the measurement time, is somewhat arbitrary.   Indeed,  separation by one standard deviation is not the only possible choice for distinguishability.  Lastly the above discussion only answers the question of how much time it takes to distinguish between two predetermine initial states of the qubit  (under a measurement with Gaussian noise). A more natural question is how long it takes the conditional state to collapse to either $z=\pm 1$. This can be characterized by the conditional variance.

\subsection {Conditional variance}

How fast the ensemble of possible trajectories converge to one of the possible values $z=\pm 1$ is characterized by the conditional variance~\cite{oxtoby:2006a}.  This measure is defined as
\begin{eqnarray} \label{eq:variance}
V_J(t) &=& \langle \op\sigma_z^2\rangle_t -\langle \op\sigma_z\rangle_t^2 =1 - z_J(t)^2.
\end{eqnarray}
It ranges from 0 to 1, with $V_J=0$ meaning that we are certain that the current value of $z_J(t)$ is either $\pm1$ and $V_J=1$ corresponding to complete uncertainty.

For $\gamma_1 =0$, Eq.~(\ref{eq:blockQTz}) can be solved analytically using \ito calculus. Doing this yields the solution
\begin{equation}
	z_J(t) = \tanh\left[ s(t) +\tanh^{-1}[z(0)] \right],
\end{equation}
where the signal $s(t)$ is defined in Eq.~(\ref{eq:signal}).  For example, in the situation where $z(0)=0$
 and arbitrary $x(0)$ and $y(0)$ (for example a completely mixed state or an $x$-eigenstate) the conditional variance becomes
\begin{equation}
	\begin{split}\label{eq:condvar}
	V_J(t) =&1 - \tanh^2\left[s(t)\right].
	\end{split}
\end{equation}

\begin{figure}[tbp]
	\centering
		\includegraphics[width=0.45\textwidth]{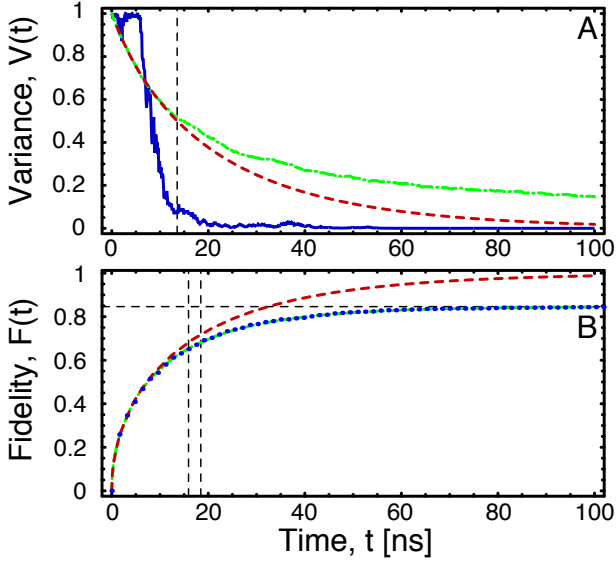}
	\caption{ (Color online) A) Typical trajectories for the conditional variance (blue solid line) with $\Gamma_\mathrm{d}/2\pi = 5$ MHz and $\eta =1$.  Average conditional variance (red dashed line) found using Eq.~\eqref{eq:condvarint} for the same parameters.  Note that the calculated ensemble average over 1000 typical trajectories falls directly over the average conditional variance, and is therefore not explicitly shown.  The vertical dash line shows the time at which the conditional variance crosses $1/2$, $t_m = 13.5$ ns.  The green dashed-dotted line is the ensemble average conditional variance for $T_1=159$ ns
($\gamma_1/2\pi = 1$ MHz) and the rest of the parameters are identical to the above.  For this value of $\gamma_1$, the SNR is $10$.
	B)  Average fidelity of the measurement at estimating an unknown but pure initial state for $\Gamma_\mathrm{d}/2\pi = 5$ MHz, $\eta =1$ and $\gamma_1/2\pi = 0$ (red dashed line) or $\gamma_1/2\pi = 1$MHz (blue dots).  The green solid line is a fit using Eq.~\eqref{eq:fidfit} with fit parameters $A_\mathrm{fit} = 0.846$ (horizontal dashed line) and $\Gamma_\mathrm{fit}/2\pi = 14.6$ MHz. The two vertical lines correspond to the times where the fidelity crosses $\mathrm{erf}(1/\sqrt{2})$.  For $\gamma_1/2\pi =0$ this yields $t_m = 15.9$ ns, while for $\gamma_1/2\pi =1$ we find $t_m=18.5$ ns}
	\label{fig:CQEDQuantumTrajectoryFig07}
\end{figure}

The conditional variance for a typical trajectory is shown in Fig.~\ref{fig:CQEDQuantumTrajectoryFig07} A) [blue solid line].  Here, we have taken $\Gamma_\mathrm{d}/2\pi = 5$ MHz and the qubit to be initially completely mixed ($x(0)=y(0)=z(0)=0$).  We see that for this typical trajectory, at about $10$ ns the measurement is completed and the state of the qubit is almost pure.  However, this is only {\em one}  typical trajectory and to get a quantitative idea of the measurement time we must consider the ensemble average of the conditional variance. This is the red dashed line in Fig.~\ref{fig:CQEDQuantumTrajectoryFig07} A).

Taking the ensemble average of Eq. \eqref{eq:condvar} gives
\begin{equation}\label{eq:condvarint}
	V(t) = \frac{1}{2\sqrt{2\pi} \nu^{3/2}}\int_{-\infty}^{\infty}\frac{x\sinh(2x) \exp(-x^2/2\nu)}{\cosh(2\nu)+\cosh(2x)}dx,
\end{equation}
where $\nu ={\Gamma_\mathrm{ci}t}$.  Using this, the measurement time $t_m$ can be defined by requiring that the average conditional variance falls below $1/2$.  Numerically, this is found to occur when $\nu = 0.851$.  Thus, using the conditional variance as the measure, the measurement time is found to be $t_m = 0.851/\Gamma_\mathrm{ci}$.

However, when $\gamma_1$ is non-zero, the conditional variance always eventually goes to zero which signifies that we are eventually certain that the system is in its ground state.  This is of course not a useful way to define the measurement time.  To take into account relaxation, in Ref.~\cite{gambetta:2007a} we studied this question by considering an effective classical model for a dispersive quantum non-demolition measurement which turns out to be equivalent to the model presented here when the initial state of the qubit is either $z(0)=\pm 1 $ or a mixture of both (i.e., $x(0)=y(0)=0$). That is, Eq.~(\ref{eq:blockQTz}) is the same as the Kushner-Stratonovich equation that describes a measurement of a two state system with Gaussian noise.  For completeness and since this is a good approach in the situation where relaxation is non-zero, we will therefore briefly review the relevant results of Ref.~\cite{gambetta:2007a}.

\subsection{Initial state fidelity}

In Ref.~\cite{gambetta:2007a}, to quantify how good a measurement is at revealing the initial state of the qubit, we imagine preparing the qubit in either the $z(0)=\pm 1$ state and then generate fictitious records $\Jp(t)$ from that input.  Given these fictitious measurement records, but now assuming ignorance of the initial state, we then ask what was the initial state of the qubit.  To quantify the efficiency of the measurement at revealing the correct input state, we use the following fidelity measure $F$: the number of correct assignments minus the number of incorrect assignments of the initial state, normalized by the total number of assignments.   This measure will range from $1$, indicating that the correct initial state was correctly found for each record, to $-1$, indicating that a wrong assigned was realized for each record.  A fidelity of $0$ implies that the assignments are completely random.

The criteria used to make these assignments is the optimal assignment criteria found in Ref.~\cite{gambetta:2007a}.  This is based on an estimate  $\tilde z(t)$ for the initial state.  If this estimate is above 0, we assign the initial state as excited and if it is below 0 we assign it as ground (for $\tilde z =0$, we randomly chose between excited or ground). This estimate $\tilde z(t)$ is defined as
\begin{equation}
	\tilde z(t) = P(z_0 = +1|\Jp,t) - P(z_0 = -1|\Jp,t),
\end{equation} where $P(z_0 = \pm1|\Jp,t) $ is the probability at time $t$ that the initial state was $\pm 1$ given the record $\Jp(t)$. This conditional probability is given by Bayes theorem
\begin{equation}
	P(z_0 |\Jp,t) = \frac{P(\Jp|z_0,t) P(z_0)}{\sum_{i=-1,1}P(\Jp|z_0=i,t) P(z_0=i)},
\end{equation} where  $P(z_0)$ is the initial state probability which we take to be 1/2 for both $i=-1$ and $1$. By introducing a fictitious unravelling of $\gamma_1$ (see Refs.~\cite{gambetta:2005a,gambetta:2007a}), the estimate can be rewritten as
 \begin{equation}
\tilde z(t) = \frac{ 1 - e^{-2s(t)}\left[1-a(t)\right]}{1+ e^{-2s(t)}\left[1+a(t) \right]},
\end{equation} where
\begin{equation}
	a(t) =\gamma_1\int_0^t d t_j e^{-\gamma_1 (t_j-t)+2s(t_j) }
\end{equation}
and $s(t)$ is the integrated signal defined in \eq{eq:signal}.

The above integral cannot be solved in general for a typical record $s(t)$, but it can be easily determined numerically. Here, however, to develop a physical understanding of this result,  we will consider the $\gamma_1=0$ limit first. Doing this, we can rewrite $\tilde z(t)$ as
\begin{equation}
	\tilde z(t) = \frac{  1 -  e^{-2s(t)} }{ 1 +  e^{-2s(t)}},
\end{equation} and using the above definition of the fidelity
\begin{equation}
	F(t) = \lim_{M\rightarrow \infty}\frac{1}{2M} \left[ \sum_{\tilde z_{+1}>0} -  \sum_{\tilde z_{+1}<0} +  \sum_{\tilde z{-1}<0} -  \sum_{\tilde z_{-1}>0}   \right]1,
\end{equation}
we obtain the following compact expression
\begin{equation}
	\begin{split}
		F(t)&=\frac{1}{\sqrt{2\pi t}} \left[\int_{-t\sqrt{\Gamma_\mathrm{ci}}}^\infty e^{-x^2/2t}dx- \int_{-\infty}^{-t\sqrt{\Gamma_\mathrm{ci}}} e^{-x^2/2t} dx\right]\\
		& =\mathrm{erf}\left(\sqrt{\frac{t \Gamma_\mathrm{ci}}{2}}\right).
	\end{split}
\end{equation}
This result is shown in Fig.~\ref{fig:CQEDQuantumTrajectoryFig07} B), for $\Gamma_\mathrm{d}/2\pi =5$ MHz,  as the red dashed line.   As can be seen from this figure, arbitrarily large fidelities can be achieved by monitoring longer.

We now use this result to define a measurement time.  A reasonable criterion is to ask for the fidelity to be at least $\mathrm{erf}(1/\sqrt{2}) = 0.68$ (corresponding to the probability of a Gaussian random variable being within one standard deviation of the mean). 
The measurement time is then defined as the time when the fidelity reaches this value.  Using the above results, we find that the measurement time is $t_\mathrm{m} = 1/\Gamma_\mathrm{ci}$.  For $\gamma_1=0$, we thus again recover the result \eq{eq:measurement_time_simple} of the simple model.

\begin{figure}[t]
	\centering
		\includegraphics[width=0.45\textwidth]{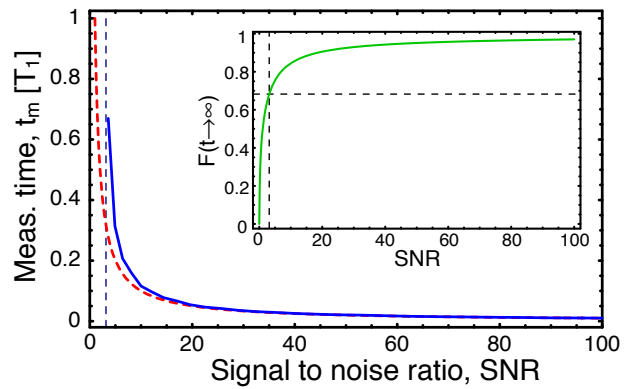}
	\caption{(Color online)  Measurement time (full blue) as extracted from the requirement that the fidelity reaches $\mathrm{erf}(1/\sqrt{2})$ as a function of the SNR ($2\Gamma_\mathrm{d}/\gamma_1$).  This is plotted for $\gamma_1/2\pi= 1$ MHz.  The vertical dashed line corresponds to SNR = $3.16$.  This defines the minimal SNR for which we can define a measurement time.  This is also presented in the inset which shows the long time fidelity as a function SNR, with the vertical dashed line being the SNR where the long time fidelity crosses $\mathrm{erf}(1/\sqrt{2})$.  The red dashed line in the main panel corresponds to $t_m = 1/\Gamma_\mathrm{ci}$ the measurement time for $\gamma_1=0$.  At low SNR, it always underestimate the measurement time. }
	\label{fig:CQEDQuantumTrajectoryFig08}
\end{figure}

The effect of relaxation on the fidelity is investigated numerically.   For example, the blue dots in Fig.~\ref{fig:CQEDQuantumTrajectoryFig07} B) show the fidelity obtained numerically for $\gamma_1/2\pi =1$ MHz.   With all values of $\Gamma_\mathrm{d}$ tried numerically, we have found excellent agreement of the fidelity, including dissipation, to the following form
\begin{equation}\label{eq:fidfit}
	F(t) = A_\mathrm{fit}\mathrm{erf}\left(\sqrt{\frac{t\Gamma_\mathrm{fit}}{2}}\right)
\end{equation} where $\Gamma_\mathrm{fit}$ and $A_\mathrm{fit}$ are fit parameters.
For example, the green solid line in Fig.~\ref{fig:CQEDQuantumTrajectoryFig07} B is for $\Gamma_\mathrm{d}/2\pi =5$ MHz). To show how the fidelity $F(t)$ depends on the SNR (which in the limit that the resonator transient can be ignored is $2\Gamma_\mathrm{d}/\gamma_1$), numerical simulations and the above described fitting procedure were performed for values of $\Gamma_\mathrm{d}/2\pi$ ranging from $1$ to $50$ MHz. The results of these simulations are shown in Fig.~\ref{fig:CQEDQuantumTrajectoryFig08}.  In particular, the inset of this figure shows the long time fidelity, $A_\mathrm{fit}$.  From this result, we see that the fidelity is below $\mathrm{erf}(1/\sqrt{2})$ until a SNR of 3.16 is reached (this corresponds to the vertical dashed line).  Below this value, using the above measurement time criteria, we must conclude that the measurement cannot be completed.  However, for all SNR above 3.16 it is possible to find a time where the fidelity crosses $\mathrm{erf}(1/\sqrt{2})$. This time is shown in Fig.~\ref{fig:CQEDQuantumTrajectoryFig08} as the blue solid line.  We see that this time is always longer then the ideal limit (red dashed line).  However, as the SNR increases, the measurement time approaches the result obtained in the $\gamma_1=0$ case ($1/\Gamma_\mathrm{ci}$).

\section{Emergence of quantum jumps}\label{sec:jumps}

By increasing the SNR (e.g. increasing the measurement drive amplitude), the conditional state will tend to be localized to either the excited or ground state.  This is due to the term $\sqrt{\Gamma_\mathrm{ci}} (1-z_J^2)$ in Eq.~(\ref{eq:blockQTz}). However, in the presence of relaxation $\gamma_1\neq0$, on average the conditional state must be localized on the ground state in the long time limit.  As a result, when initializing the  system in the excited state, there are two competing processes.  One is trying to lower the conditional state to $z=-1$ ($\gamma_1$) and the second is trying to hold it in at $z=+1$ ($\Gamma_\mathrm{ci}$).  The net result is that as the measurement strength becomes stronger, the SME will transform from a diffusive stochastic process which stochastically lowers itself to the ground state to one which jumps at a well defined time.  That is, as the measurement becomes stronger,  quantum jumps due to the decay channel $\gamma_1$ will be revealed in the signal $J(t)$.

\begin{figure}[tbp]
	\centering
		\includegraphics[width=0.45\textwidth]{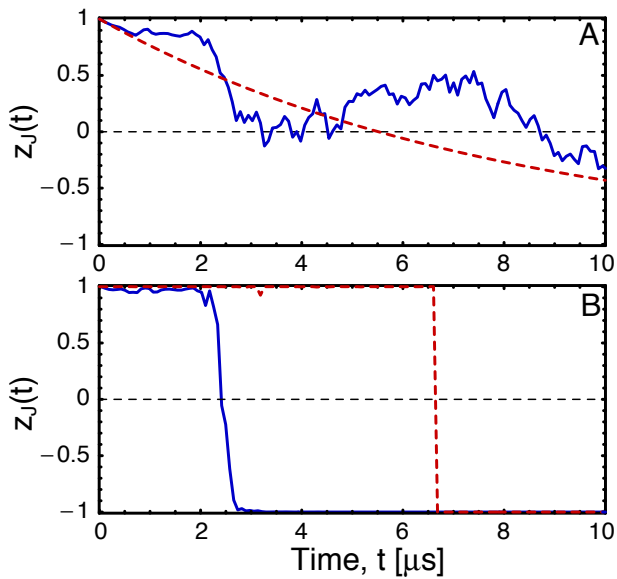}
	\caption{(Color online) Time evolution of the $z$ component of the  conditional state for a qubit initially in the excited state and $\chi/2\pi =5$ MHz, $\kappa/2\pi = 10$ MHz, $\Delta_r=0$, $\eta =1$ and $T_1= 7\:\mu$s. A) Measurement drive strength of $\varepsilon_\mathrm{d}/2\pi = 0 $ MHz (red dashed) and  0.2 MHz (solid blue).  B)  $\varepsilon_\mathrm{d}/2\pi = 0.9$ MHz (solid blue) and 20 MHz (dashed red).}
	\label{fig:CQEDQuantumTrajectoryFig09}
\end{figure}

The crossover of the conditional state from a diffusive to jump-like SME is numerically shown in Fig.~\ref{fig:CQEDQuantumTrajectoryFig09} for measurement drive strengths $\varepsilon_\mathrm{d}/2\pi = \{ 0, 0.2,0.9,20\}$ MHz, respectively.  From this figure, we see that when $\varepsilon_\mathrm{d}/2\pi = 0$ (dashed red in panel A), as expected,  the state simply decays to the ground state at a rate  $\gamma_1$ as this is equivalent to gaining no information about the qubit state.  For $\varepsilon_\mathrm{d}/2\pi =  0.2$ MHz (full blue in panel A) it stochastically follows this decay curve.  However,  at $\varepsilon_\mathrm{d}/2\pi =0.9$ MHz (full blue in panel B), there appears a sharp transition from the excited state to the ground state.  As the measurement amplitude is further increased ($\varepsilon_\mathrm{d}/2\pi = 20$ MHz, red dashed in panel B), this transition becomes sharper and happens at a well defined time.

\begin{figure}[tbp]
	\centering
		\includegraphics[width=0.45\textwidth]{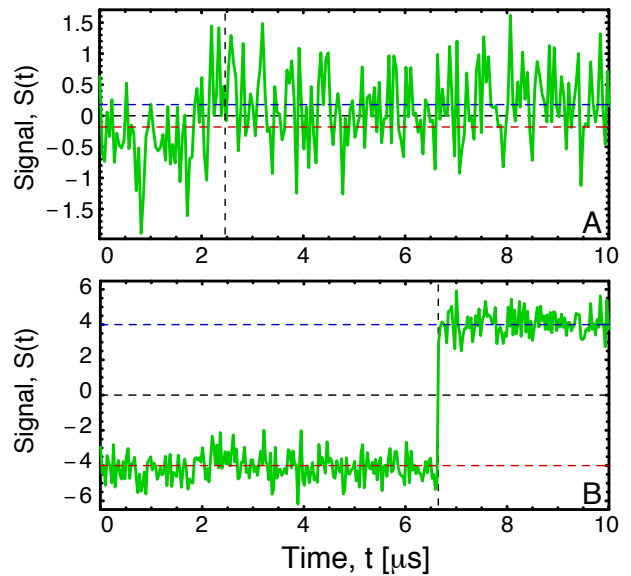}
	\caption{(Color online) Integrated signal $S(t)$ as a function of time.  A) Measurement drive strength $\varepsilon_\mathrm{d}/2\pi =0.9$ MHz and integration steps $\delta t = 48$ ns  B) $\varepsilon_\mathrm{d}/2\pi =20$ MHz and $\delta t = 32$ ns. The vertical dashed line indicates the time where the conditional state crosses 0 in Fig. \ref{fig:CQEDQuantumTrajectoryFig09}.  The two horizontal dashed lines are the expected results corresponding to $z=\pm1$ constant.}
	\label{fig:CQEDQuantumTrajectoryFig10}
\end{figure}

Experimentally, one does not have direct access to $z_J(t)$ but reconstructs this quantity from the measured current $J(t)$.  As a result, a more striking demonstration of quantum jumps would be to see these jumps directly in $J(t)$.  However, due to white noise fluctuations $\xi(t)$, without first averaging the signal over some finite integration time the jumps are not visible in most trajectories.  As a result, we consider the quantity
\begin{equation} \label{eq:ss}
S(t) = \frac{1}{\delta t} \int_t^{t+\delta t} \frac{ J(s)}{\sqrt{\kappa\eta}}ds
\end{equation}
where $\delta t$ is an adjustable integration parameter.  At large $\delta t$, the fluctuations are well suppressed by the averaging, but at the expense of smoothing out the jump.  For example, Fig.~\ref{fig:CQEDQuantumTrajectoryFig10} shows the integrated current corresponding to the two measurement amplitudes used in Fig.~\ref{fig:CQEDQuantumTrajectoryFig09} B).   Panel A) of Fig.~\ref{fig:CQEDQuantumTrajectoryFig10} corresponds to $\varepsilon_\mathrm{d}/2\pi =0.9$ MHz and $\delta t$ = $48$ ns, while panel B) corresponds to $\varepsilon_\mathrm{d}/2\pi =20$ MHz and $\delta t$ = $32$ ns.  From this figure, we see that for $\varepsilon_\mathrm{d}/2\pi =0.9$ MHz one cannot resolve the jump from the integrated signal.   However, by increasing the drive to $\varepsilon_\mathrm{d}/2\pi =20$ MHz the jump becomes clearly visible.

Thus we see that if we wish to see the jump in the integrated signal we need a much higher signal to noise ratio. This is because by definition the conditional state is the best way we can process the record. We note that it should be possible to optimize the processing of the current to distinguish the jump at lower SNR by using a more complicated filter then time averaging.  For example, one could easy envisage that there exists an optimal integration time $\delta t = t_\mathrm{opt}$, or kernel $k(t)$, such that the integrated signal, $S(t) = \int_{-\infty}^\infty k(t,s)J(s)dt$, would do almost as good as the conditional state (see Ref.~\cite{gambetta:2007a} where a similar question was investigated).

Finally, an interesting situation occurs if the $\gamma_1$ channel can be directed  (corresponding to the fluorescence of the ``atom''). When we register that the qubit has jumped, we know that the energy has been dissipated into the $\gamma_1$ channel. If this fluorescence is directed to an additional low-Q resonator, this can be used as a single microwave photon source with a well-defined emission time. Inversely, if the qubit is prepared in the ground state this second resonator can be used as a single photon detector. When a jumped of the qubit is registered, we know that a photon was present in the second cavity. The efficiency of this detector will be limited by the Zeno effect~\footnote{A. Blais, J. Gambetta, C. Cheung, R. Schoelkopf, and S. M. Girvin, manuscript in preparation}.

\section{Weak Driving of the Qubit -- Zeno Effect} \label{sec:zeno}

In the previous sections, we considered the dynamics of the qubit under measurement.  In this section, we will include the possibility of having a control drive.  As explained in Sec.~\ref{sec:circuit_QED}, this is done in practice by adding a drive on the input port of the resonator at the qubit transition frequency.   Our goal is to use the transformation~\eq{Uframe} to again obtain an effective SME for the qubit only.   However, as shown in appendix~\ref{sec:Control}, in the transformed frame, the drive on the qubit induces transitions between the different (transformed) states of the resonator.  As a result, it is not possible to integrate out exactly the resonator from the effective description of the qubit.

However, in the small $\beta$ limit, we find we find that qubit evolution is well described by the SME {
\begin{equation}\label{eq:Rabi}
\begin{split}
\dot{\qrho}_\Jp(t)=& -i  \frac{\Omega_R}{2}\left[\sx,\qrho_\Jp(t)\right]
	+\left[\gamma_{\phi} + \Gamma_\mathrm{d}(t)\right]{\cal D} [\sz] \qrho_\Jp(t)/2\\&-i \frac{\sqrt{\Gamma_\mathrm{ba}}}{2} [\szo,\qrho_\Jp(t)] (\Jp(t) - \sqrt{\Gamma_\mathrm{ci}}\langle \op\sigma_z \rangle_t)\\&+\sqrt{\Gamma_\mathrm{ci}}\Mso{\szo}{\langle\szo\rangle}\qrho_\Jp(t) (\Jp(t)
- \sqrt{\Gamma_\mathrm{ci}}\langle \op\sigma_z \rangle_t)\\&+	\gamma_1{\cal D}[\smm] \qrho_\Jp(t)
\end{split}
\end{equation}
{where $\Omega_R$ is the Rabi frequency of the control drive tuned to the qubit frequency, $ \omega_\mathrm{ac}$.}  The competition between Rabi flopping induced by the control drive and the extra dephasing due to the measurement drive should enable us to observe the Zeno effect in this system.  Indeed, as the measurement rate $\Gamma_\mathrm{ci}$ is increased, the qubit dynamics will switch from Rabi oscillations to jump-like behavior.  In the limit of very large measurement rates,  all dynamics will disappear and the qubit will remain fixed in the $z_J = \pm 1$ state (if $\gamma_1$ is zero).

To see this more clearly, the above SME can be written in terms of the components of the Bloch vectors as
\begin{equation}
	\begin{split}
		\dot{x}_J(t)=& -\{\gamma_2+\Gamma_\mathrm{d}(t) +\sqrt{\Gamma_\mathrm{ci}} z_J(t)[\Jp(t)-\sqrt{\Gamma_\mathrm{ci}}z_J(t)]\}  x_J(t)  \\
		        \dot{y}_J(t)=&-\Omega_R z_J(t) -\{\gamma_2+\Gamma_\mathrm{d}(t) \\&+\sqrt{\Gamma_\mathrm{ci}} z_J(t)[\Jp(t)-\sqrt{\Gamma_\mathrm{ci}}z_J(t)]\}  y_J(t)\\
		         \dot{z}_J(t)=&\Omega_R y_J(t)-\gamma_1[z_J(t)+1]\\&+ \sqrt{\Gamma_\mathrm{ci}} [1-z_J(t)^2] [\Jp(t) - \sqrt{\Gamma_\mathrm{ci}}z_J(t)].
	\end{split}
\end{equation}
 Here we have assumed that $\phi=\theta_\beta$ such that $\Gamma_\mathrm{ba}=0$.
In the limit that the measurement induced dephasing rate $\Gamma_\mathrm{d}$ is much larger than all other rates, we can set $\dot x_J(t) = \dot y_J(t) = 0$ to arrive at
\begin{equation}
	\begin{split}
			\dot z_J(t) =& -2\gamma_\mathrm{jump}z_{J}(t)-\gamma_1(z_{ J}(t) + 1) \\&+ \sqrt{\Gamma_\mathrm{ci}}(1-z_{ J}(t)^2)(\Jp(t) -\sqrt{\Gamma_\mathrm{ci}}z_{ J}(t)).
	\end{split}
\end{equation}
As expected, in this expression, the jump rate is given by $\gamma_\mathrm{jump}=\Omega_R^2/2(\gamma_2+\Gamma_\mathrm{d})$.  From this expression, we can indeed expect Zeno-type dynamics for the qubit. { Note that the stochastic term $\sqrt{\Gamma_\mathrm{ci}}z_J(t)[\Jp(t)-\sqrt{\Gamma_\mathrm{ci}}z_J(t)]$ has been ignored in the denominator of the jump rate.}

\begin{figure}[t]
	\centering
		\includegraphics[width=0.45\textwidth]{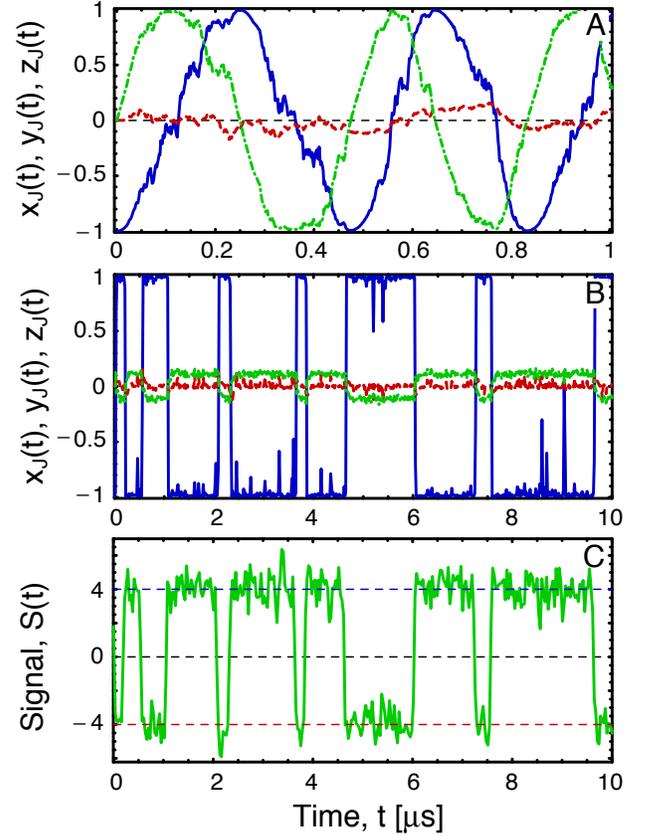}
	\caption{(Color online) Time evolution of the conditional state for the full homodyne SME for a qubit initially in the ground state: $x$ red dashed, $y$ green dashed-dotted and $z$  solid blue. The system parameters are $\chi/2\pi =5$ MHz, $\kappa/2\pi = 10$ MHz, $\Delta_r=0$, $\eta =1$, $\gamma_1 = \gamma_\phi =0$ and $\Omega_\mathrm{R}/2\pi = 2.5$ MHz.  A) Measurement drive amplitude of $\varepsilon_\mathrm{d}/2\pi = 0.9$ MHz.  B)  $\varepsilon_\mathrm{d}/2\pi = 20$  MHz. C) Integrated signal with $\delta t =32$ ns and for a measurement amplitude of $\varepsilon_\mathrm{d}/2\pi = 20$ MHz.}\label{fig:CQEDQuantumTrajectoryFig11}
\end{figure}

\begin{figure}[t]
	\centering
		\includegraphics[width=0.45\textwidth]{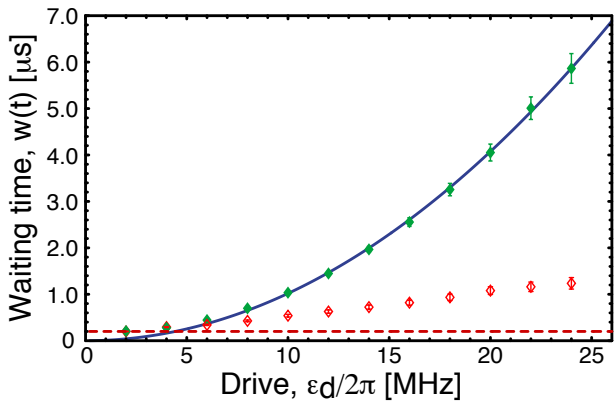}
	\caption{(Color online) Average jump time as a function of the measurement drive strength $\varepsilon_d$. The jump time is defined as the time for the system state to go from $z=\pm 0.95$ to $z=\mp 0.95$. The red dashed line corresponds to the Rabi period $\pi/\Omega_\mathrm{R}$ of 200 ns for a Rabi drive of $\Omega_\mathrm{R}/2\pi=2.5$ MHz. The blue solid line corresponds to the expected mean jump time expected from the Zeno effect, $\gamma_\mathrm{jump}=\Omega_R^2/2(\gamma_2+\Gamma_\mathrm{d})$.  This agrees very well with the numerically simulations of Eq. \eqref{eq:Rabi} (green filled diamonds) where the mean jump time was extracted from a total simulation time of 16 ms.  The red open diamonds are the mean jump time as extracted from the full numerical homodyne SME for a total simulation time of  about 15 ms.}
	\label{fig:CQEDQuantumTrajectoryFig12}
\end{figure}

As discussed in Sec.~\ref{sec:jumps}, as the measurement becomes stronger (i.e. $\Gamma_\mathrm{ci}$ increases) the conditional state will lock itself into either the excited or ground state ($z=\pm 1$).  Because of the
control drive, whose amplitude enters in the jump rate $\gamma_\mathrm{jump}$, random fluctuations will result in the system jumping between these two localized states.    However, unlike the situation presented in the previous section, the rate of the these jumps also depends on the measurement strength through $\Gamma_\mathrm{ci}$.  As a result, in the limit that the measurement SNR is large, the jumps will disappear and the qubit will remain fixed in either $z = \pm 1$.  This is the Zeno effect \cite{breuer:2002a,Gagen:1993a,Presilla:1996a,Cresser:2006a}.

To illustrate this, we have numerically integrated the full homodyne SME (\ref{eq:HomodyneQT}), with the addition of a qubit drive, for 2 different measurement} drives $\varepsilon_d/2\pi = 0.9$ MHz and  20 MHz, and a Rabi frequency of $\Omega_R/2\pi= 2.5$ MHz. The results are shown in Fig.~\ref{fig:CQEDQuantumTrajectoryFig11}.  From this figure, we see that for the low measurement amplitude (pannel A), measurement only causes small amplitude noise on the Rabi oscillations.  However, for the larger amplitude (panel B) the conditional dynamics is jump-like.  We note that the rate of the jumps between the ground and excited state is, as expected from the above discussion, slower then the rate $\Omega_R$ at which the Rabi drive would coherently drive the system. Note that the x-axis scale is not the same in pannel A and B(C).

These jumps can be observed directly in the signal $J(t)$.  This is illustrated in Fig. \ref{fig:CQEDQuantumTrajectoryFig11} C) for the case of the larger measurement amplitude.  As in the previous section, we are showing here the integrated signal, defined in Eq. \eqref{eq:ss}, and with a
integration time $\delta t = 32$ ns.  The integrated signal clearly reveals the jumps.

We note that the numerical results presented here were obtained using the full homodyne SME, and not the simplified SME~\eqref{eq:Rabi}. This is because while this simplified equation contains the essential features, it is not strictly valid in the limit of large mean photon number.  {However, by inspecting individual trajectories, we find that trajectories obtained from the full model display the same qualitative behavior as those obtained from the effective model.  The difference between the two types of trajectories appear to be only small changes in renormalized system parameters.}

{To explore the difference between these two models in more detail, we simulated both models as a function of the measurement drive strength $\varepsilon_\mathrm{d}$.} The average jump time extracted for both cases is shown in Fig. \ref{fig:CQEDQuantumTrajectoryFig12}.  The simple model (green filled diamonds) agrees with the predicted jump time $1/\gamma_\mathrm{jump}$ for large $\varepsilon_\mathrm{d}$ (blue solid line).  In the low measurement drive limit, it agrees with the time set by the Rabi drive (red dashed line).  However, the full homodyne SME calculations (red open diamonds) do not agree with this expectation in the large drive case.  Although the mean jump time does increase with the measurement drive as expected by the Zeno effect, its effect is not as large as that predicted by the simple model. .


\section{Conclusion} \label{sec:con}

We have investigated the measurement of a superconducting qubit using a dispersively coupled resonator.  From the fully quantum mechanical model for the resonator and qubit dispersive interaction, we have found an effective master equation for the qubit (only) under the presence of a measurement drive on the resonator.  This was done by moving to a  frame which takes into account the entanglement between the resonator and the qubit.  With respect to the bare qubit-resonator master equation, the effective qubit master equation contains an additional,  time-dependent decay channel which has the form of a dephasing process and which depends on the number of photons populating the resonator.  This is referred to as measurement-induced dephasing~\cite{devoret:2000a,averin:2002a,clerk:2003a,makhlin:2001a,gambetta:2006a}.  In addition to the extra dephasing channel, the photon population of the resonator is also responsible for an ac-Stark shift of the qubit transition frequency.  As was shown in Refs.~\cite{dykman:1987a,gambetta:2006a}, and later reported experimentally in Ref.~\cite{schuster:2007a}, these effects lead to {\em number splitting} of the qubit spectrum in the limit where $\chi\gg \kappa$.  We thus have obtained a very simple to solve (2 dimensional Hilbert space) effective model for the qubit dynamics that contains the essential physics.  We believe that this is an interesting analytical, and numerical, tool to further explore the dispersive regime.

We then focused on the situation where one is continuously monitoring the voltage at the output port of the resonator.  In that situation, the conditional state of the combined system evolves according to the homodyne SME~\cite{wiseman:1993a}.  Again moving to a frame that takes into account the qubit-resonator entanglement, we then obtained an effective SME for the qubit that is valid in the limit $\gamma_1\ll \kappa$.  The SME corresponds to a weak measurement of the qubit operator $\op\sigma_z$.  As a result, the measurement stochastically projects the conditional state into eigenstates of $\op\sigma_z$.  This is similar to the result that was obtained in Refs.~\cite{korotkov:1999b,korotkov:2001a,korotkov:2001b,korotkov:2003a,wiseman:2001b,goan:2001a,oxtoby:2005a,oxtoby:2007a} for a double quantum dot qubit monitored continuously by a quantum point contact (although the results are presented in very different forms in all of these references).

From these results, we then showed how, in the limit that the measurement becomes strong, there is a cross-over in the qualitative behavior of the trajectories from diffusive to jump-like.  This process relies on the $T_1$ decay channel for the qubit to jump from the excited to the ground state.  As a result, we expect to see quantum jumps, with Poisson statistics, in the measured homodyne current.

Moreover, in presence of a Rabi drive on the qubit, we again saw that the trajectories evolve from diffusive to jump-like.  In this situation and when the measurement drive is not too strong, the jump rate is given $\Omega_R^2/2(\gamma_2+\Gamma_\mathrm{d})$.  This is due to a competition between the measurement drive that tends to lock the conditional state on one of the eigenstates of $\op\sigma_z$ and the Rabi drive that tends to flip the qubit.  This is nothing but the Zeno effect.  However, as the measurement gets stronger, the simple dispersive Hamiltonian approximation breaks down and 
the effective model fails to predict the correct mean jump time for the qubit.  We found that the form of the trajectories are still consistant with the model, only the parameters appear to be renormalized by the stronger measurement drive.  An interpretation of this discrepancy is that the resonator response is not instantaneous.
When the qubit jumps, the resonator state will lag behind and as a result the effective separation in phase space between the two coherent states of the resonator corresponding to the two qubit states will be reduced.  In turn, this can lead to a reduction of the measurement induced-dephasing.

Overall, the model obtained here can be used to obtain qualitative information about the qubit dynamics in the presence of dissipation, measurement and control.  We also gave a numerical prescription to go beyond this simple model.

\begin{acknowledgments}
We thank Jens Koch for discussions. SMG, JG, AAH, and DIS were supported by the NSA and LPS under ARO W911NF-05-1-0365 and by the NSF under DMR-0653377 and DMR-0603369. AB was supported by the Natural Sciences and Engineering Research Council of Canada (NSERC), the Fond Qu\'eb\'ecois de la Recherche sur la Nature et les Technologies (FQRNT) and the Canadian Institute for Advanced Research (CIFAR).  MB was supported by the Natural Sciences and Engineering Research Council of Canada (NSERC).
\end{acknowledgments}

\appendix

\section{Derivation of the effective qubit master equation}
\label{frame}


In this appendix we show how to obtain the effective qubit master equation \eq{eq:MasterEqQubit} from the full dispersive  qubit-resonator master equation \eq{eq:MasterEq}.  For this purpose, we introduce the the transformation
\begin{equation} \label{EQ.appendix.Uframe}
\tP = \op\Pi_e\op{D}(\alpha_e)+\op\Pi_g\op{D}(\alpha_g),
\end{equation}
where $\op\Pi_{e,g}$ are the projectors on the qubit excited and ground state and $\op D(\alpha)$ is the
field displacement operator~\cite{walls:1994a}.

In this frame, the state of the combined qubit-resonator system $\crho\trans{\tP}=\tP^\dag\crho\tP$ can be written in the energy basis as
\begin{equation}  \label{R.energy}
\crho\trans{\tP} = \sum_{n,m =0}^\infty \, \sum_{i,j =e,g} \crho\trans{\tP}_{n,m,i,j} \ket{n,i}\bra{m,j}.
\end{equation}
Using this expression for the system's density matrix in the transformed frame, it is simple to express the qubit's reduced density matrix in laboratory frame as
\begin{equation} \label{pho.energy}
\begin{split}
\qrho &= {\rm Tr}_{\rm res.}[ \tP \crho\trans{\tP} \tP^\dag ] \\
&= \sum_n [\crho\trans{\tP}_{n,n,e,e} \ket{e}\bra{e} +\crho\trans{\tP}_{n,n,g,g} \ket{g}\bra{g} ] \\
&+ \sum_{n,m} [\lambda_{n,m,m,n} \ket{e}\bra{g} + \lambda^*_{m,n,n,m} \ket{g}\bra{e}]
\end{split}
\end{equation}
with
\begin{eqnarray}\label{eq:definitionlambda}
\lambda_{n,m,p,q} = \crho\trans{\tP}_{n,m,e,g}d_{p,q} \exp\left[-i{\rm Im}[\alpha_g\alpha_e^*]\right]
\end{eqnarray} and $d_{p,q}$ is the matrix element of the displacement operator in the photon number basis
\begin{eqnarray}
d_{p,q} =\bra{p} \op D (\beta) \ket{q}.
\end{eqnarray}

In the remainder of this appendix, we show how by using the transformed master equation for $\crho\trans{\tP}$ it is possible to obtain the coefficients entering in the expression~\eq{pho.energy} for $\qrho$.

\subsection{The displaced frame}

In the displaced frame, the state matrix $\crho\trans{\tP}$ obeys the master equation
\begin{equation}
\begin{split}
    \dot{\crho}\trans{\tP}= &
    -\frac{i}{\hbar}[\op{H}\trans{\tP}_{\rm eff},\crho\trans{\tP}]
    +\kappa{\cal D}[\op{a}\trans{\tP}]\crho\trans{\tP}
    +\gamma_1{\cal D}[\op{\sigma}\trans{\tP}_-]\crho\trans{\tP}
    +\gamma_{\phi}{\cal  D}[\op{\sigma}\trans{\tP}_z]\crho\trans{\tP}/2\\
    &
    - \tP^\dag\dot \tP\crho\trans{\tP} - \crho\trans{\tP} \dot \tP^\dag \tP
\end{split}
\label{Eq.Transformed.ME}
\end{equation}
with $\op O\trans{\tP} = \tP^\dag \op O\tP$ the transformed operators and where the effective Hamiltonian $\op{H}_{\rm eff}$ is given by~\eq{eq:FreeHam3}.  Using the standard results
\begin{equation}
\begin{split}
\op{D}\dg(\alpha)\op{a}\op{D}(\alpha) &= \aop+\alpha \\
\op{D}\dg(\alpha)\op{a}\dg\op{D}(\alpha) &= \ad+\alpha^*
\end{split}
\end{equation}
we obtain
\begin{equation}
\begin{split}
a\trans{\tP} &= \aop  +\Pa\\
(a^\dag a)\trans{\tP} &=  \ad\aop + \ad\Pa + \aop\Pa^* + |\agg|^2\Pg+ |\aee|^2\Pe
\end{split}
\end{equation}
where we have defined
\begin{equation}
\Pa = \agg\Pg + \aee\Pe.
\end{equation}
For the qubit operators, we obtain
\begin{equation}
\begin{split}
\sz\trans{\tP} &=  \sz\\
\sigma_-\trans{\tP} &=  \smm D^\dag(\agg)D(\aee).
\end{split}
\label{EQ.Qubit.transform}
\end{equation}
Using these results, we then have for the transformed Hamiltonian\begin{equation}
\begin{split}
\op H\trans{\tP}_{\rm eff}
&=\frac{\hbar\tilde\omega_a}{2}\op{\sigma}_z
+ \hbar \left[\veps_{d}(\ad + \Pa^*)+\mathrm{h.c.}\right]\\
&+\hbar(\Delta_r+\chi\sz)
 \left(
 \ad\aop +\ad\Pa+\aop\Pa^*+|\Pa|^2
 \right)
\end{split}
\end{equation}

We now turn to the damping superoperators.  For the field damping, we find
\begin{equation}
\begin{split}
	\D{\op a\trans{\tP}}\crho\trans{\tP} =&
	\D{\aop}\crho\trans{\tP} - \frac12 [\ad\Pa-a\Pa^*,\crho\trans{\tP}] \\
	& + \frac{\beta^*}{2}a[\crho\trans{\tP},\sz] + \frac{\beta}{2}[\sz,\crho\trans{\tP}]\ad \\
	& + \frac{\Gamma_\mathrm{m}(t)}{4\kappa} \D{\sz}\crho\trans{\tP} -i \frac{{\rm Im}[\alpha_g\alpha_e^*]}{2}[\sz,\crho\trans{\tP}],
\end{split}
\end{equation}
where
\begin{equation}
\Gamma_\mathrm{m}  = \kappa  {|\beta|^2}
\end{equation}
and
\begin{equation}
\beta  = \aee - \agg.
\end{equation}
In this frame, damping of the field induces extra dephasing of the qubit at a rate $\Gamma_\mathrm{m}$.
Note this is not the measurement induced dephasing rate given in Ref.~\cite{gambetta:2006a}, as the present
expressions are in the transformed frame, not the laboratory frame.
Transforming the qubit dissipation superoperators, we obtain
\begin{align}
\D{\op{\sigma}_z\trans{\tP}}\crho\trans{\tP}  &= \D{\sz}\crho\trans{\tP},\\
\D{\op{\sigma}_-\trans{\tP}}\crho\trans{\tP}  &= \D{\smm D^\dag(\agg)D(\aee)}\crho\trans{\tP}.
\end{align}

The last terms to take into account are those in the last line of \eq{Eq.Transformed.ME}.
For these terms, we use the fact that
\begin{equation}
\dot D(\alpha)
=
\left[
(\dot\alpha \ad - \dot\alpha^*\aop) +(\dot\alpha^*\alpha-\dot\alpha\alpha^*)/2
\right]
D(\alpha)
\end{equation}
to obtain
\begin{equation}
\begin{split}
- \tP^\dag\dot \tP\crho\trans{\tP} - \crho\trans{\tP} \dot \tP^\dag \tP =& -\com{\dPa\ad - \dPa^*\aop}{\crho\trans{\tP}} \\
& -i\com{{\rm Im}[\daee\aee^*]\Pi_e + {\rm Im}[\dagg\agg^*]\Pi_g}{\crho\trans{\tP}}.
\end{split}
\end{equation}

Choosing $\daee$ and $\dagg$ in the above expression as in \erf{eq:alpha}, we finally obtain the
transformed master equation
\begin{equation}
\begin{split}
\dot{\crho}\trans{\tP} =
&
-i \frac{\tilde\omega_a+\tilde B}{2}\left[\sz,\crho\trans{\tP}\right]
-i \left[(\Delta_r+\chi\sz)\ad\aop,\crho\trans{\tP}\right]\\
&
+\gamma_1 \D{\smm D^\dag(\agg)D(\aee)}\crho\trans{\tP}
+\left[\gamma_{\phi} + \Gamma_\mathrm{m}/2 \right]\D{\sz} \crho\trans{\tP}/2\\
&
+\kappa\D{\aop}\crho\trans{\tP}
+\frac{\kappa\beta}{2}\left[\sz,\crho\trans{\tP}\right]\ad
+\frac{\kappa\beta^*}{2}\aop\left[\crho\trans{\tP},\sz\right],
\label{EQ.R.Transformed.final} 
\end{split}
\end{equation}
where we have defined the time-dependent shifted qubit transition frequency as
\begin{equation}
\tilde B=(\beta \veps_d^*+\beta^* \veps_d)/2 +\kappa{\rm Im}[\alpha_g\alpha_e^*].
\end{equation}
In this transformed frame, the last two terms of \eq{EQ.R.Transformed.final} appear because the resonator acts as a non-Markovian bath for the qubit~\cite{imamoglu:1994a,gambetta:2002a}.  Moreover, we note that the term proportional to $\gamma_1$ in \eq{EQ.R.Transformed.final} can be expressed as
\begin{equation}
\begin{split}
&\D{\smm D^\dag(\agg)D(\aee)}\crho\trans{\tP} \\
&\qquad
=  \smm D(\beta) \crho\trans{\tP} \spp D^\dag(\beta)-\spp\smm\crho\trans{\tP}/2
-\crho\trans{\tP}\spp\smm/2.
\end{split}
\label{EQ.D.tranformed.sm}
\end{equation}
Upon taking the trace over the resonator space, this expression will simply take the form of regular Markovian $\gamma_1$ damping on the qubit.  This is one of the features of the transformed master equation that will allow us below to obtain a simple effective master equation for the qubit only.

\subsection{Moving back to the laboratory frame}

Using \eq{R.energy}  and \eq{EQ.R.Transformed.final}, we get the following set of coupled differential equations for the coefficients of the laboratory frame reduced qubit density matrix~\eq{pho.energy}
\begin{widetext}
\begin{align}
\label{eq:matrixelementdiffequations1}
\dot{\crho}\trans{\tP}_{n,m,e,e}
=& [-i\Delta_r(n-m) - i \chi(n-m) - \gamma_1 - \kappa(n+m)/2]\crho\trans{\tP}_{n,m,e,e} + \kappa \crho\trans{\tP}_{n+1,m+1,e,e} \sqrt{(n+1)(m+1)} \\
\label{eq:matrixelementdiffequations2}
\begin{split}
\dot{\crho}\trans{\tP}_{n,m,g,g}
=& [-i\Delta_r(n-m) + i \chi(n-m)  - \kappa(n+m)/2]\crho\trans{\tP}_{n,m,g,g} + \gamma_1 \sum_{p,q} \crho\trans{\tP}_{p,q,e,e} d_{n,p}d_{m,q}^* \\
&+ \kappa \crho\trans{\tP}_{n+1,m+1,g,g} \sqrt{(n+1)(m+1)}
\end{split}
\\
\label{eq:matrixelementdiffequations3-.5}
\dot \lambda_{n,m,p,q} =& \dot \crho\trans{\tP}_{n,m,e,g} d_{p,q} e^{-i{\rm Im}[\agg\aee^*]} - i\partial_t({\rm Im}[\agg\aee^*])\lambda_{n,m,p,q} + \dot\beta\sqrt{p}\lambda_{n,m,p-1,q} - \dot\beta^*\sqrt{q}\lambda_{n,m,p,q-1} - \frac12\partial_t(\beta\beta^*)\lambda_{n,m,p,q}
\\
\label{eq:matrixelementdiffequations3}
\begin{split}
=& [-i (\tilde\omega_a + B) -i\Delta_r(n-m) - i \chi(n+m) - \gamma_1/2 - \gamma_\phi -\Gamma_\mathrm{d} - \kappa(n+m)/2] \lambda_{n,m,p,q}\\
& + \kappa\lambda_{n+1,m+1,p,q} \sqrt{(n+1)(m+1)} +\kappa \beta\sqrt{m+1} \lambda_{n,m+1,p,q} - \kappa \beta^*\sqrt{n+1} \lambda_{n+1,m,p,q} + \dot{\beta} \sqrt{p} \lambda_{n,m,p-1,q}\\
&  - \dot{\beta}^* \sqrt{q} \lambda_{n,m,p,q-1}
\end{split}
\end{align}
\end{widetext}
where \eqref{eq:matrixelementdiffequations3-.5} was obtained by differentiating \eqref{eq:definitionlambda}, with respect to time and
\begin{equation}
\Gamma_\mathrm{d}  = 2\chi  {\rm Im}[\alpha_g \alpha_e^*]
\end{equation}
and
\begin{equation}
B = 2 \chi {\rm Re}[\alpha_g \alpha_e^*].
\end{equation}

To reconstruct the effective master equation from these amplitudes, we note that the time-derivative of the $\ket{e}\bra{e}$ component of \eq{pho.energy} can be written as
\begin{equation}\label{eq:rhoee}
\dot \qrho_{e,e}
=  \sum_n \dot{\crho}\trans{\tP}_{n,n,e,e} = -\gamma_1 \sum_n  \crho\trans{\tP}_{n,n,e,e} =-\gamma_1\qrho_{e,e}.
\end{equation} For the ground state component we have 
\begin{align}
\label{eq:rhogg}
\dot\qrho_{g,g} &= \sum_n \dot{\crho}\trans{\tP}_{n,n,g,g} =\gamma_1 \sum_{n,p,q} {\crho\trans{\tP}}_{p,q,e,e} d_{n,p}d_{n,q}^* = \gamma_1\qrho_{e,e}.
\end{align}
where in the last equality we have used the identity
\begin{equation}
\sum_{n=0}^\infty d_{n,p} d^*_{n,q} = \delta_{ pq}.
\end{equation}
The off diagonal terms are the hardest to deal with, but because in the transformed frame, the photon population is initially zero and there is no mechanism to populate  $\lambda_{n,m,p,q}$ terms with $\{n,m,p,q\}>0$ we have
\begin{align}
\label{eq:rhoeg}
\dot\qrho_{e,g} &= \dot \lambda_{0,0,0,0} = [-i (\tilde\omega_a + B) - \gamma_1/2 - \gamma_\phi -\Gamma_\mathrm{d}] \qrho_{e,g}.
\end{align}

The above expressions correspond to the following effective master equation for the qubit
\begin{equation}
\begin{split}
\dot\qrho
= &
 -i  \frac{\omega_\mathrm{ac}}{2}\left[\sz,\qrho\right]+\gamma_1 \D{\smm}\qrho
+\left[\gamma_{\phi} + \Gamma_\mathrm{d} \right]\D{\sz} \qrho/2\\
\equiv &  \mathcal{L} \qrho,
\end{split}
\label{EQ.reduced.rho}
\end{equation}
where $\omega_\mathrm{ac}=\tilde\omega_a+B$.


\section{Derivation of the effective qubit stochastic master equation}\label{app:qt}

To derive the effective qubit stochastic master equation we start by using the linear form \cite{gambetta:2005a,wiseman:1996a,goetsch:1994a} of the full SME~\erf{eq:HomodyneQT}. This is 
\begin{equation}
	\begin{split}
		\dot{\bar\crho}_J=&{\cal
		L}_{\rm tot}\bar\crho_J+\sqrt{\kappa\eta}\bar{\cal M}[2\op I_\phi]\bar\crho_J J + i\sqrt{\kappa\eta}[\op Q_\phi,\bar\crho_J] J.
	\end{split}
\end{equation}
where the bar here is used to signify that the state is not normalized and the linear measurement superoperator is
\begin{equation}
	\bar{\cal M}[\op c]\crho = \op c \crho/2 + \crho\op c/2.
\end{equation}
Moving to the frame defined by \erf{Uframe} gives
\begin{equation} \label{eq:linear_trajectory_full}
	\begin{split}
		\dot{\bar\crho}\trans{\tP}_J=&{\cal
		L}_{\rm tot}{\bar\crho\trans{\tP}}_J+\sqrt{\kappa\eta} \Big{[}|\beta|\cos (\theta_\beta-\phi)\bar{\cal M}[\szo]{\bar\crho\trans{\tP}}_J \\& +|\mu|\cos(\theta_\mu -\phi){\bar\crho\trans{\tP}}_J+ \ano e^{-i\phi}{\bar\crho\trans{\tP}}_J + {\bar\crho\trans{\tP}}_J\cro e^{i\phi} \Big{]} J \\&i\frac{\sqrt{\kappa\eta}|\beta|\sin(\theta_\beta-\phi)}{2}[\szo,{\bar\crho\trans{\tP}}]J.
	\end{split}
\end{equation}
In the same way as \erft{eq:matrixelementdiffequations1}{eq:matrixelementdiffequations3} were obtained, we can write equations of motion for the coefficients ${\bar \crho\trans{\tP}}_{n,m,e,e}$, ${\bar \crho\trans{\tP}}_{n,m,g,g}$ and $\bar\lambda_{n,m,p,q}$ of the decomposition of the conditional density matrix ${\bar\crho\trans{\tP}}_J$ in the energy basis.  The first term of \eq{eq:linear_trajectory_full} yields the same terms as in \erft{eq:matrixelementdiffequations1}{eq:matrixelementdiffequations3}.  As a result, we get
\begin{widetext}
\begin{align}
\label{eq:matrixelementdiffequationsqt1}
\begin{split}
\dot{\bar\crho}\trans{\tP}_{n,m,e,e}
= & (\ref{eq:matrixelementdiffequations1}) + \sqrt{\kappa\eta}[\sqrt{n+1} e^{-i\phi}{{\bar \crho\trans{\tP}}}_{n+1,m,e,e}+\sqrt{m+1} e^{i\phi}{{\bar \crho\trans{\tP}}}_{n,m+1,e,e} + |\mu|\cos(\theta_\mu -\phi){{\bar \crho\trans{\tP}}}_{n,m,e,e}\\
& + |\beta|\cos(\theta_\beta -\phi){{\bar \crho\trans{\tP}}}_{n,m,e,e} ]J
\end{split}\\
\label{eq:matrixelementdiffequationsqt2}
\begin{split}
\dot{\bar\crho}\trans{\tP}_{n,m,g,g}
= & (\ref{eq:matrixelementdiffequations2}) + \sqrt{\kappa\eta}[\sqrt{n+1} e^{-i\phi}{{\bar \crho\trans{\tP}}}_{n+1,m,g,g}+\sqrt{m+1} e^{i\phi}{{\bar \crho\trans{\tP}}}_{n,m+1,g,g} + |\mu|\cos(\theta_\mu -\phi){{\bar \crho\trans{\tP}}}_{n,m,g,g}\\
&- |\beta|\cos(\theta_\beta -\phi){{\bar \crho\trans{\tP}}}_{n,m,g,g} ]J
\end{split}\\
\label{eq:matrixelementdiffequationsqt3}
\begin{split}
\dot{\bar\lambda}_{n,m,p,q}
= & (\ref{eq:matrixelementdiffequations3}) +  \sqrt{\kappa\eta}[\sqrt{n+1} e^{-i\phi}{{\bar \crho\trans{\tP}}}_{n+1,m,e,g}d_{p,q}e^{-i\mathrm{Im}[\alpha_g\alpha_e^*]}+\sqrt{m+1} e^{i\phi}{{\bar \crho\trans{\tP}}}_{n,m+1,e,g}d_{p,q}e^{-i\mathrm{Im}[\alpha_g\alpha_e^*]} \\
& + |\mu|\cos(\theta_\mu -\phi) {{\bar\lambda}}_{n,m,p,q} +i |\beta|\sin(\theta_\beta -\phi) {{\bar\lambda}}_{n,m,p,q} ]J.
\end{split}
\end{align}
\end{widetext}
In the above expressions, the equation numbers refer to the RHS of the corresponding expressions.  They are the contribution of the Lindblad term ${\cal L}_{\rm tot}{\bar\crho\trans{\tP}}_J$.

By noting that these matrix elements are not coupled in the qubit basis,  we can use the same arguments that were used to obtain  \erf{eq:rhoeg}.   Doing this, we find for the stochastic parts
\begin{align}
\begin{split}
\dot{\bar\qrho}_{e,e}
& = \dot{\bar\crho}\trans{\tP}_{0,0,e,e} \\
& = \sqrt{\kappa\eta}\Big{[}|\mu|\cos(\theta_\mu -\phi)+ |\beta|\cos(\theta_\beta -\phi) \Big{]}\times\bar\qrho_{e,e} J
\end{split}\\
\begin{split}
\dot{\bar\qrho}_{e,g}
& = \dot{\bar\lambda}_{0,0,0,0}\\
& = \sqrt{\kappa\eta}\Big{[}|\mu|\cos(\theta_\mu -\phi)+ i|\beta|\sin(\theta_\beta -\phi) \Big{]}\times\bar\qrho_{e,g} J.
\end{split}
\end{align}
For the ground state amplitude we have
\begin{equation}
\begin{split} \label{eq:b9}
\dot{\bar\qrho}_{g,g}
= & \sum_n \dot{\bar\crho}\trans{\tP}_{n,n,g,g}\\
= &\sqrt{\kappa\eta}\Big{[}|\mu|\cos(\theta_\mu -\phi)- |\beta|\cos(\theta_\beta -\phi) \Big{]}\\
&\times \bar\qrho_{g,g} J+\sqrt{\kappa\eta}\Big{[}(e^{-i\phi}\bar\crho\trans{\tP}_{1, l}+ e^{i\phi}\bar\crho\trans{\tP}_{1, r}) \Big{]}J.
\end{split}
\end{equation}
where we have defined
\begin{align}\label{eq:b10}
\bar \crho\trans{\tP}_{1, l} &= \sum_{n}\sqrt{n+1} \, \bar\crho\trans{\tP}_{n+1,n,g,g}\\
\label{eq:b11}
\bar \crho\trans{\tP}_{1, r} &= \sum_{n}\sqrt{n+1} \, \bar\crho\trans{\tP}_{n,n+1,g,g}.
\end{align}
The amplitude $\bar\crho\trans{\tP}_{1, l}$ obeys the equations of motion
\begin{equation}
\begin{split}
\dot{\bar \crho}\trans{\tP}_{1, l}
=
& -i (\Delta_r-\chi) \bar\crho\trans{\tP}_{1, l}-\kappa\bar \crho\trans{\tP}_{1, l}/2   +\beta\gamma_1  \bar \qrho\trans{\tP}_{e,e} \\
& + \sqrt{\kappa\eta}\Big{[}|\mu|\cos(\theta_\mu -\phi)- |\beta|\cos(\theta_\beta -\phi) \Big{]}\bar \crho\trans{\tP}_{1, l}J  \\
& + \sqrt{\kappa\eta}[\sum_{n}\sqrt{(n+2)(n+1)} \bar\crho\trans{\tP}_{n+2,n,g,g}e^{-i\phi}  \\
& + \sum_{n}n \bar\crho\trans{\tP}_{n,n,g,g}e^{i\phi}]J
\end{split}
\end{equation}
and we find a similar equation for $d_t \bar \crho\trans{\tP}_{1, r}$.  From this result, we find that provided that
\begin{equation}
\epsilon \equiv \frac{2\gamma_1}{\kappa}\ll 1,
\end{equation}
the effect of  $\bar\crho\trans{\tP}_{1, l} $ and $\bar\crho\trans{\tP}_{1, r}$ on $\bar\qrho\trans{\tP}_{g,g}$  will be very small.  In this situation,  it is possible to construct a laboratory frame linear SME for the qubit only:
\begin{equation}\label{eq:linearqubitQT}
\begin{split}
\dot{\bar\qrho}_{ J}
=
& {\cal L}\bar \qrho_{ J}+\sqrt{\kappa\eta}|\beta|\cos(\theta_\beta-\phi)\bar{\cal M}[\szo]\bar\qrho_{ J} J \\
& +i \frac{\sqrt{\kappa\eta}|\beta|\sin(\theta_\beta-\phi)}{2} [\szo, \bar\qrho_{J}]  J \\
&+ \sqrt{\kappa\eta}|\mu|\cos(\theta_\mu -\phi)\bar\qrho_{ J} J.
\end{split}
\end{equation}
Using \erf{eq:relation} and normalizing the above SME gives \erf{eq:qubitQT} with measurement record statistics given by \erf{eq:normcurrent}.


\section{Including coherent control of the qubit}
\label{sec:Control}

A drive on the input port of the resonator at a frequency $\omega_{c}$ close to the qubit transition frequency can be used to coherently control the state of the qubit.  As shown in Refs.~\cite{blais:2004a,blais:2007a}, this can be represented by the term
\begin{equation}
\frac{\Omega_R}{2}\left(  \smm  e^{+i \omega_{c} t}  +  \spp e^{-i \omega_{c} t}  \right),
\end{equation}
in the qubit--resonator Hamiltonian \eq{eq:FreeHam3}.  In this expression, $\Omega_R$ is the Rabi amplitude which depends on the drive amplitude and its detuning to the resonator frequency.

To see how this changes the effective qubit master equations obtained in the previous appendices, we apply the transformation of \eq{Uframe}.  In a frame rotating at the frequency frequency $\omega_{c}$, we find
\begin{equation}
\frac{\Omega_R}{2}\left(e^{-i \mathrm{Im}\left\{\agg\aee^*\right\}} \smm D(\beta) + {\rm h.c.}\right).
\end{equation}
In this transformed frame, it is apparent that when the qubit flips, it must `drag' the photon field populating the resonator.    In this situation, the equation of motion for the coefficients of \eq{pho.energy} are modified in the following way
\begin{align}
\dot\crho\trans{\tP}_{n,m,e,e}
& =  (\ref{eq:matrixelementdiffequations1})
+ i \frac{\Omega_R}{2} \sum_p [\lambda_{n,p,p,m}-\lambda^*_{m,p,p,n}] \\
\dot\crho\trans{\tP}_{n,m,g,g}
& = (\ref{eq:matrixelementdiffequations2})
-i \frac{\Omega_R}{2} \sum_p [\lambda_{p,n,m,p}-\lambda^*_{p,n,m,p}] \\
\dot \lambda_{n,m,p,q}
& = (\ref{eq:matrixelementdiffequations3})
-i \frac{\Omega_R}{2} \sum_l [ \crho\trans{\tP}_{l,m,g,g} d^*_{l,n}-\crho\trans{\tP}_{n,l,e,e} d^*_{m,l}]d_{p,q},
\end{align}
where the equation numbers refer to the RHS of the corresponding expressions.  These additional terms mix the various coefficients of the decomposition of $\qrho$ and obtaining an effective master equation for the qubit only is no longer possible exactly.  This can nevertheless be done in the small measurement amplitude limit.


The case where a control and a measurement drive are acting on the system simultaneously is most interesting in the case where the measurement drive is of small amplitude.  For example, in Ref.~\cite{wallraff:2005a}, a weak continuous measurement drive corresponding to about one photon was used to monitor Rabi oscillations.   When the measurement drive amplitude is increased, the qubit is dephased faster and coherent control is realized with less fidelity.

In the weak measurement limit $\beta \rightarrow 0$, the matrix element of the displacement operator $d_{n,m}$ in the
photon number basis can be approximated to
\begin{equation}
\begin{split}
 d_{p,q}
& = e^{-|\beta|^2/2} \sqrt{\frac{{\rm min}(p,q)!}{{\rm max}(p,q)!}}  \\
& \times  L_{{\rm min}(p,q)}^{{\rm abs}(p-q)} (|\beta|^2)
\begin{cases}
\beta^{p-q},  		&	p>q\\
(-\beta^*)^{q-p},	&	q>p
\end{cases}
\\
& \approx e^{-|\beta|^2/2} \sqrt{\frac{{\rm min}(p,q)!}{{\rm max}(p,q)!}} \left[1 - \left({\rm max}(p,q)+1/2 \right) |\beta|^2\right]  \\
&
\quad
\times
\begin{cases}
\beta^{p-q},  		&	p>q\\
(-\beta^*)^{q-p},	&	q>p
\end{cases},
\end{split}
\end{equation}
where $L_n^{m} (x)$ is an associated Laguerre polynomial.  To lowest order in $\beta$, this reduces $ d_{p,q} \sim \delta_{p,q}$.  Making this replacement in the above expressions for the amplitudes, we find in this limit the following effective qubit master equation
\begin{equation}
\begin{split}
\dot\qrho
= &
 -i  \frac{\tilde\omega_a+B}{2}\left[\sz,\qrho\right] -i  \frac{\Omega_R}{2}\left[\sx,\qrho\right]\\
&
+\gamma_1 \D{\smm}\qrho
+\left[\gamma_{\phi} + \Gamma_{\mathrm{d}} \right]\D{\sz} \qrho/2.
\end{split}
\label{EQ.reduced.rho.drive}
\end{equation}
Unsurprisingly, in the small $\beta$ limit, the only effect of a control drive is to flip the qubit at the Rabi frequency $\Omega_R$. If we were to include the next order in $d_{p,q}$ we would not be able to obtain an equation for just the qubit, effects such as sidebands would be observed.


\end{document}